\documentclass[aps,prd,nofootinbib,longbibliography,superscriptaddress]{revtex4-2}

\usepackage{graphicx}
\usepackage{amsmath,amssymb,mathtools,bm}
\usepackage{booktabs}
\usepackage{array}
\usepackage{multirow}
\usepackage{tabularx}
\usepackage{microtype}
\usepackage{comment}
\usepackage{xcolor}
\usepackage[colorlinks=true,linkcolor=blue,citecolor=blue,urlcolor=blue]{hyperref}
\graphicspath{{./}{figures/}}

\newcommand{\SNR}{\mathrm{SNR}}
\newcommand{\SNRc}{\mathrm{SNR}_{C}}
\newcommand{\SNRr}{\mathrm{SNR}_{R}}
\newcommand{\AU}{\mathrm{AU}}
\newcommand{\pc}{\mathrm{pc}}
\newcommand{\m}{\mathrm{m}}
\newcommand{\km}{\mathrm{km}}
\newcommand{\s}{\mathrm{s}}
\newcommand{\um}{\mu\mathrm{m}}
\newcommand{\uas}{\mu\mathrm{as}}
\newcommand{\mas}{\mathrm{mas}}
\newcommand{\nas}{\mathrm{nas}}

\newcommand{\FRC}{\mathrm{FRC}}
\newcommand{\SSIM}{\mathrm{SSIM}}
\newcommand{\SBR}{\mathrm{SBR}}

\newcolumntype{Y}{>{\raggedright\arraybackslash}X}

\begin{document}

\title{Ultra-High-Resolution Astronomy with the Solar Gravitational Lens}

\author{Slava G. Turyshev}
\affiliation{Jet Propulsion Laboratory, California Institute of Technology,\\
4800 Oak Grove Drive, Pasadena, CA 91109-0899, USA}

\date{\today}

\begin{abstract}
The solar gravitational lens (SGL) is a target-specific physical-optics observatory: the Sun supplies the dominant wave-optical element, while a spacecraft or formation in the focal region supplies occultation, calibrated annular photometry, image-plane sampling, metrology, and inverse reconstruction.  We develop a quantitative observability framework for representative non-exoplanet SGL astronomy, not an end-to-end mission validation.  Target viability is controlled by the source-to-image mapping $\boldsymbol{\rho}=-(z/z_0)\boldsymbol{\xi}$, image-plane diameter $D_{\rm img}=z\Theta$, raster pitch $\Delta_{\rm img}=D_{\rm img}/n$, finite-source gain scale $\mu_{\rm ext}D_{\rm img}\simeq4b$, source-to-background ratio, temporal coherence, PSF knowledge, calibration covariance, metrology, and focal-line access.  We distinguish the full vector Poisson measurement operator from the stationary aperture-averaged scalar convolution used for controlled benchmarks. Four analytic scenes are propagated and reconstructed: a solar analog and magnetic white dwarf at $10\,\pc$, an M87*-scale compact millimeter ring/jet source, and a bright $0.1\,\AU$ protoplanetary subfield at $140\,\pc$.  Under stated kernel-mismatch, background, calibration-floor, support-mask, sampling, and regularization assumptions, and an imposed effective convolved-raster information floor $\SNRc$, the scalar reconstructions give $\SSIM=0.993$, $0.918$, $0.973$, and $0.923$.  These metrics quantify scalar inverse conditioning, not delivered flight performance; truth-referenced $\FRC_{50}$, support-leakage, and $\SSIM$-versus-$\SNRc$ diagnostics make the dependence on the assumed information floor explicit. Many self-luminous compact targets are not photon-starved relative to a reflected-light exo-Earth reference, so the dominant requirements become calibrated ring extraction, solar/coronal subtraction, detector dynamic range, PSF knowledge, cadence, spectroscopy, metrology, scan overhead,  and focal-line access. The strongest bounded astronomical cases are white-dwarf surface and magnetic mapping, nearby stellar surfaces, compact AGN/black-hole structure with dedicated long-wavelength instrumentation, velocity-resolved broad-line-region mapping, and selected planet-forming subfields.  A separate highest-priority enabling program is SGL transfer-function characterization: measuring the solar-multipole, plasma, extended-Sun, and instrumental components of the SGL response needed to make such imaging scientifically interpretable.
\end{abstract}

\maketitle
\tableofcontents

\section{Introduction}
\label{sec:intro}

The Solar Gravitational Lens (SGL) offers an observational regime that cannot be reached by any practical conventional telescope or engineered interferometer \cite{Turyshev2025DirectHighRes,Turyshev2026DirectHighResExo}.  Diffraction of electromagnetic waves in the Sun's gravitational field produces a high-gain interference region beginning at the heliocentric distance $R_\odot^2/(2r_g)\simeq548\,\AU$, where $r_g=2GM_\odot/c^2$ is the solar Schwarzschild radius \cite{Einstein1936,Eshleman1979,Turyshev2017PRD95,TuryshevToth2017PRD96}.  The SGL does not form a conventional focal-plane image: an extended source is mapped into a parity-reversed image cylinder in the focal region, so a spacecraft-borne observing system must suppress the solar disk, measure the annular Einstein-ring signal around the occulted Sun, sample the image plane, and solve an inverse problem \cite{TuryshevToth2020Photometric,TuryshevToth2020Process,TuryshevToth2020Extended,TothTuryshev2021Recovery}.

Resolved exoplanet imaging has been the main driver for detailed SGL studies because the combination of faint reflected light, angular scale, and scientific value is compelling \cite{TuryshevToth2022Resolved,TuryshevToth2022Spectral,TothTuryshev2023Rotating,Turyshev2026DirectHighResExo}.  The same physical optics, however, open a broader astronomical question: which non-exoplanet targets are compact, bright, slow enough or modelable enough, and unique enough to justify a preselected SGL focal line?  The answer is not obtained from point-source gain or formal angular resolution alone.  Large fields project to impractical image-plane sizes, faint diffuse structures lose finite-source gain, rapidly evolving sources require dynamic inversion, and retargeting by even $1^\circ$ requires a transverse displacement of order $10\,\AU$ at $650\,\AU$.

This paper fills that gap by developing a physics-of-observability framework for representative non-exoplanet SGL targets.  The target set is not an exhaustive survey of all astrophysics or cosmology; it is a stress test of classes for which public literature already defines angular scales, brightness scales, variability, and observables.  The portfolio includes stellar surfaces and magnetic activity \cite{CHARA2005,Monnier2007,Roettenbacher2016,Berdyugina2005,DonatiLandstreet2009}, magnetic white dwarfs \cite{WickramasingheFerrario2000,Ferrario2015}, black-hole and AGN structure \cite{EHT2019I,EHT2019IV,EHT2022SgrA,Gravity2018BLR,GravityAGN2022,BlandfordMcKee1982,Peterson1993}, protoplanetary subfields \cite{ALMAPartnership2015,Andrews2018}, compact strong-lens and microlensed sources \cite{DalalKochanek2002,VegettiKoopmans2009,Gilman2020,Wambsganss2006,Kochanek2004}, and selected Local Group or high-redshift compact sources \cite{Riess2022}.  The contribution is a traceable chain from target scale to finite-source gain, photon/background estimates, scalar inverse conditioning, robustness diagnostics, and qualitative priority.

The central thesis is that the SGL should be treated as a constrained gravitational observatory.  Its governing variables are not only the formal angular response and ideal gain, but also the vector source-to-image mapping, finite-source gain, scan cost, and source dynamics:
\begin{equation}
  \boldsymbol{\rho}=-z\boldsymbol{\theta}
  =-\frac{z}{z_0}\boldsymbol{\xi},\qquad
  D_{\rm img}=z\Theta,\qquad
  \Delta_{\rm img}=\frac{D_{\rm img}}{n},\qquad
  N=n^2,\qquad
  \Delta s=\frac{z_0}{z}\Delta_{\rm img},\qquad
  \chi_t=\frac{T_{\rm scan}}{t_{\rm var}} .
\label{eq:governing_variables}
\end{equation}
Here $z$ is the heliocentric receiver distance, $z_0$ is the target distance, $\boldsymbol{\xi}$ is a source-plane coordinate, $\boldsymbol{\theta}=\boldsymbol{\xi}/z_0$ is the angular coordinate, $\boldsymbol{\rho}$ is the SGL image-plane coordinate, $\Theta$ is the angular diameter of the target or selected subfield, $D_{\rm img}$ is the image-plane diameter, $\Delta_{\rm img}$ is the image-plane raster pitch, $n$ is the linear raster dimension, $N=n^2$ is the total number of image-plane samples, and $t_{\rm var}$ is the source variability time.  A strong SGL target is compact but not point-like, photon-rich or calibratable, temporally reconstructable, and scientifically unique relative to foreseeable alternatives. Table~\ref{tab:notation} summarizes notations and abbreviations used throughout. 

\paragraph*{Scope of this paper.}
This paper defines observability closure for representative non-exoplanet SGL targets.  The analysis has three goals: identify target classes for which the SGL provides unique spatial or spatial-spectral information; quantify the image-plane, radiometric, temporal, and calibration scales that make those targets viable; and translate scalar reconstruction benchmarks into explicit technology requirements.  The simulations use a stationary scalar aperture-averaged SGL response, analytic source scenes, prescribed metric supports, imposed effective $\SNRc$ floors, and fixed regularization.  These choices isolate target selection and inverse conditioning.  Full observatory validation requires the corresponding annular/vector measurement operator, wavelength dependence, solar-multipole and plasma transfer functions, detector covariance, metrology, and dynamic source models.

The evidentiary chain is explicit: compute $D_{\rm img}$, $\Delta_{\rm img}$, $\mu_{\rm ext}$, $Q_s/Q_b$, and $\chi_t$; test scalar recovery with fixed kernels, masks, noise floors, and metrics; then rank applications by uniqueness and closure risk.  The present results establish geometric/radiometric closure and controlled scalar recoverability.  The verification path from these results to target-specific observatory performance is correspondingly specific: build the physical SGL transfer-function library, propagate annular/vector data with detector and occulter models, close detector/background covariance, demonstrate image-plane metrology at the required fraction of $\Delta_{\rm img}$, and perform dynamic Bayesian reconstructions for variable sources.

We proceed as follows.  Section~\ref{sec:optical_model} presents the SGL optical and observing model.  Section~\ref{sec:phase_space} defines target phase space and observability metrics.  Section~\ref{sec:simulation_method} describes radiometric and scalar reconstruction benchmarks.  Section~\ref{sec:sim_results} presents the reconstruction and robustness results.  Section~\ref{sec:spectroscopy} discusses spatially resolved spectroscopy and spectropolarimetry.  Section~\ref{sec:science_cases} evaluates science cases, Sec.~\ref{sec:comparison} compares the SGL with foreseeable observatories, Sec.~\ref{sec:mission} derives mission and observing requirements, Sec.~\ref{sec:ranking} prioritizes applications, Sec.~\ref{sec:limitations} summarizes limitations, and Appendix~\ref{app:benchmark_details} summarizes the benchmark implementation details used for the scalar simulations.

\begin{table}[t!]
\centering
\caption{Notation and abbreviations.  The raster convention follows the resolved exoplanet-imaging benchmark: $n$ is the linear raster dimension and $N=n^2$ is the total number of image-plane samples.}
\label{tab:notation}
\begin{tabular}{ll}
\toprule
Symbol or abbreviation & Meaning \\
\midrule
SGL & solar gravitational lens \\
PSF & point-spread function \\
SNR & signal-to-noise ratio \\
SBR & source-to-background ratio, $\SBR=Q_s/Q_b$ \\
FRC & Fourier-ring correlation; $\FRC_{50}$ is used here as a truth-referenced benchmark resolution diagnostic \\
FOV & field of view, used only for conventional instruments or selected SGL subfields \\
AU, pc & astronomical unit and parsec \\
$z$, $z_0$ & heliocentric receiver distance and target distance \\
$z_{\rm foc}$ & solar-grazing focal onset, $R_\odot^2/(2r_g)$ \\
$\boldsymbol{\rho}$, $\boldsymbol{\xi}$ & SGL image-plane and source-plane coordinates \\
$D_{\rm img}$ & projected image-plane diameter of the target or selected subfield \\
$n$, $N$ & linear raster dimension and total sample count, $N=n^2$ \\
$\Delta_{\rm img}$ & image-plane pitch, $D_{\rm img}/n$ \\
$d$ & telescope diameter \\
$K(\rho)$ & aperture-averaged scalar SGL kernel \\
$Q_{\rm exo}$, $Q_{\rm cor}$ & reflected exo-Earth source and solar-corona rates used as the exoplanet reference \\
$Q_s$, $Q_b$ & generalized source and residual-background rates for arbitrary target classes \\
$\SNRc$, $\SNRr$ & convolved-raster SNR and post-reconstruction residual SNR statistic \\
$\sigma_{\rm det}$ & per-sample detector-noise term, Eq.~(\ref{eq:snrc_general}) \\
$\epsilon_{\rm cal}$, $\epsilon_{\rm sys}$ & multiplicative background-calibration and systematic source residual fractions \\
$C_{\rm rec}/C_{\rm true}$ & percentile-based contrast recovery on mask $M$, Eq.~(\ref{eq:contrast_def}) \\
$f_{\rm out}$ & out-of-support leakage fraction for compact-mask cases, Eq.~(\ref{eq:fout_def}) \\
$f_{\rm smear}$ & allowed motion smear as a fraction of $\Delta_{\rm img}$, Eq.~(\ref{eq:smear}) \\
$\chi_t$ & temporal-coherence ratio, $T_{\rm scan}/t_{\rm var}$ \\
$f_J$ & unmodeled fraction of the low-order multipole displacement \\
${\cal G}_{\rm eff}$ & end-to-end effective photon-coupling gain, Eq.~(\ref{eq:geff}) \\
\bottomrule
\end{tabular}
\end{table}

\section{SGL observing geometry and optical response}
\label{sec:optical_model}

\subsection{Focal-line geometry and vector source--image mapping}

The SGL strong-interference region begins when solar-grazing rays reach the optical axis \cite{Eshleman1979,Turyshev2017PRD95,TuryshevToth2017PRD96},
\begin{equation}
z_{\rm foc}\simeq \frac{R_\odot^2}{2r_g}\simeq547.8\,\AU .
\end{equation}
At heliocentric distance $z$, the dominant impact parameter and angular Einstein-ring radius are
\begin{equation}
  b(z)=\sqrt{2r_g z},\qquad
  \theta_E(z)=\frac{b}{z}=\sqrt{\frac{2r_g}{z}} .
  \label{eq:btheta}
\end{equation}
Eqs.~(\ref{eq:btheta}) are the same geometric scales used in SGL image-formation studies \cite{TuryshevToth2020Photometric,TuryshevToth2020Extended}. At $650\,\AU$, $b=7.58\times10^8\,\m=1.089R_\odot$ and the ring-limb clearance is only $\theta_E-R_\odot/z\simeq0.132''$, so solar suppression and coronal calibration are first-order requirements.  Moving outward increases $b$ and the ring-limb clearance but increases cruise time, communication range, and retargeting distance.  We therefore restrict focal-region plots to $548$--$1000\,\AU$; the practical optical/NIR operating interval considered here is approximately $650$--$900\,\AU$, with $1000\,\AU$ treated as an upper reference rather than a baseline.

For a distant source at target distance $z_0$, the vector mapping follows from the small-angle image-cylinder geometry used in extended-source SGL imaging.  A transverse source-plane coordinate $\boldsymbol{\xi}$, an angular coordinate $\boldsymbol{\theta}=\boldsymbol{\xi}/z_0$, and an SGL image-plane coordinate $\boldsymbol{\rho}$ are related by the parity-reversing mapping
\begin{equation}
  \boldsymbol{\rho}=-z\boldsymbol{\theta}
  =-\frac{z}{z_0}\boldsymbol{\xi}.
  \label{eq:vector_mapping}
\end{equation}
The minus sign is a convention-dependent image inversion about the optical axis, but it is physically useful: an eastward displacement on the source maps to a westward displacement in the image plane.  All scalar diameter, sampling, and radiometric estimates use the norm of Eq.~(\ref{eq:vector_mapping}).  Thus
\begin{equation}
  D_{\rm img}=|\boldsymbol{\rho}_{\max}-\boldsymbol{\rho}_{\min}|=z\Theta,
  \label{eq:dimg}
\end{equation}
where $\Theta$ is the angular diameter of the target or selected subfield.  For a source of physical diameter $L_{\rm src}$,
\begin{equation}
  D_{\rm img}=L_{\rm src}\frac{z}{z_0},\qquad
  \Delta s=\Delta_{\rm img}\frac{z_0}{z}
  =317\,\m\left(\frac{\Delta_{\rm img}}{1\,\m}\right)
  \left(\frac{z_0}{1\,\pc}\right)
  \left(\frac{650\,\AU}{z}\right),
  \label{eq:source_sampling}
\end{equation}
where $\Delta s$ is the corresponding source-plane displacement.  Thus source-plane sampling is extremely fine for nearby stellar targets but becomes AU-scale for cosmological AGN.

The scalar simulations in this paper implement Eq.~(\ref{eq:vector_mapping}) by treating the simulated source as an image-plane brightness field $O_{\rm img}(\boldsymbol{\rho})=O_{\rm src}(-z_0\boldsymbol{\rho}/z)$.  The sign changes orientation but not convolution conditioning, radiometry, or scalar reconstruction metrics; source-plane products are returned in the original source orientation by applying the inverse parity transform.

\begin{figure*}[t]
  \centering
  \includegraphics[width=0.86\textwidth]{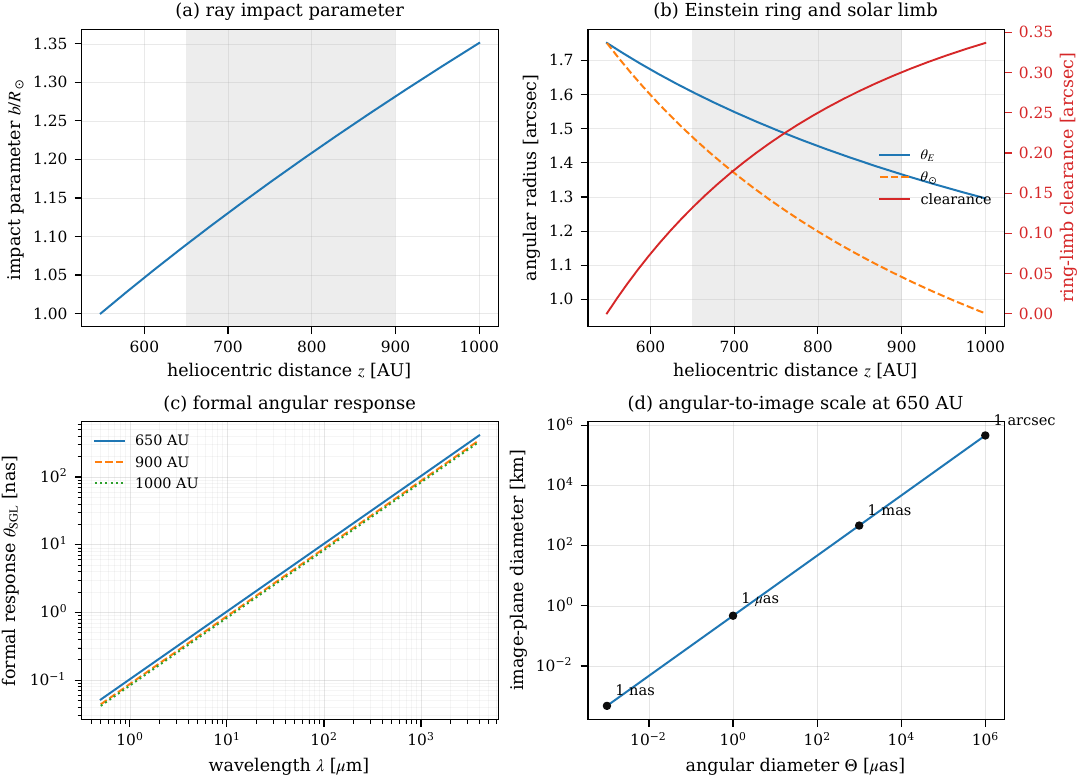}
  \caption{SGL observatory scalings restricted to $548$--$1000\,\AU$.  Panel (a) gives the ray impact parameter $b(z)$ in solar radii.  Panel (b) compares the Einstein-ring radius, solar angular radius, and ring-limb clearance.  Panel (c) gives the formal angular response $\theta_{\rm SGL}=0.38\lambda/b$ for $650$, $900$, and $1000\,\AU$.  Panel (d) converts angular diameter to image-plane diameter at $650\,\AU$.  The shaded region marks the nominal $650$--$900\,\AU$ optical/NIR operating range.}
  \label{fig:scalings}
\end{figure*}

At $650\,\AU$, the angular-to-image conversions are
\begin{align}
  1\,\nas &\rightarrow 0.471\,\m,\nonumber\\
  1\,\uas &\rightarrow 0.471\,\km,\nonumber\\
  1\,\mas &\rightarrow 471\,\km,\nonumber\\
  1'' &\rightarrow 4.71\times10^5\,\km .
  \label{eq:mapping_numbers}
\end{align}
Target access is equally restrictive in magnitude:
\begin{equation}
  |\Delta \boldsymbol{\rho}_{\rm retarget}|=z|\Delta\boldsymbol{\alpha}|
  =11.3\,\AU\left(\frac{z}{650\,\AU}\right)
    \left(\frac{\Delta\alpha}{1^\circ}\right).
  \label{eq:retarget}
\end{equation}
The SGL is therefore a preselected-line-of-sight instrument, not a survey telescope. Figure~\ref{fig:scalings} summarizes these focal-line, ring-limb, formal-resolution, and angular-to-image-plane scalings over the $548$--$1000\,\AU$ range.

\subsection{Finite-source gain and source-plane sampling}

The point-source gain $\mu_0$ is not the appropriate signal scale for a resolved target; finite-source averaging is the controlling radiometric effect for extended SGL images \cite{TuryshevToth2020Extended,TuryshevToth2022Resolved}.   Let $\theta_R=R/z_0$ be the angular radius of a compact approximately circular source, so that $\Theta=2\theta_R$ and $D_{\rm img}=2z\theta_R$. The finite-source average over the monopole response scales as
\begin{equation}
  \mu_{\rm ext}\simeq \min\left[\mu_0,\frac{2\theta_E}{\theta_R}\right]
  =\min\left[\mu_0,\frac{4b}{D_{\rm img}}\right].
  \label{eq:mu_ext}
\end{equation}
The second equality follows directly from $\theta_E=b/z$ and $D_{\rm img}=2z\theta_R$; it is therefore valid for the geometry used here, independent of the parity sign in Eq.~(\ref{eq:vector_mapping}).  In the resolved-source limit it gives
\begin{equation}
  \mu_{\rm ext}D_{\rm img}\simeq4b
  =3.03\times10^9\,\m\left(\frac{z}{650\,\AU}\right)^{1/2}.
  \label{eq:mu_ext_product}
\end{equation}
This relation is a finite-source scaling, not a replacement for a full annular throughput calculation.  It assumes a compact source or selected subfield whose integrated flux is averaged over the SGL response.  For a highly structured extended object, the correct calculation is the operator integral in Eq.~(\ref{eq:vector_measurement}); Eq.~(\ref{eq:mu_ext_product}) should then be applied to the selected field or local component being scanned, not blindly to an entire galaxy or disk.

The effective photon-coupling gain in a real instrument is bounded by finite-source averaging, aperture averaging, ring extraction, occulting throughput, detector coupling, and calibration losses:

\begin{equation}
{\cal G}_{\rm eff}(\lambda,z,d,{\cal I}_{\rm inst},D_{\rm img})
=
\eta_{\rm ann}(\lambda,{\cal I}_{\rm inst})
\min\left[
\bar{\mu}_d(d,\lambda,z),
\frac{4b}{D_{\rm img}},
\mu_0
\right],
\label{eq:geff}
\end{equation}
where $\bar{\mu}_d$ is the aperture-averaged point-source gain, $\eta_{\rm ann}$ is an end-to-end annular throughput, and ${\cal I}_{\rm inst}$ denotes the instrument, occulting, detector-coupling, and extraction configuration.  The equality sign in Eq.~(\ref{eq:geff}) defines the effective gain used for the benchmark radiometric estimates; a detailed instrument model would replace $\eta_{\rm ann}$ with an annular extraction and detector-coupling calculation.  This is the independent check on the rate equation below: a target-independent gain is not valid unless it is explicitly interpreted as an already computed ${\cal G}_{\rm eff}$ for that target class.

Eqs.~(\ref{eq:vector_mapping})--(\ref{eq:geff}) are the key checks preventing the misuse of point-source gain for large, low-surface-brightness targets.  They also show why extended targets are not forbidden.  A large object can be scanned across the image plane to build a map and a spectral data cube; the cost is that $D_{\rm img}$, scan time, and finite-source gain all scale against the observer for whole-field imaging.  The SGL is therefore best used for compact targets, selected subfields, or sparse mapping of extended systems.

\subsection{Formal angular response and equivalent baseline}

For a monopole Sun, the wave-optical point-source amplification contains the Bessel factor \cite{Turyshev2017PRD95,TuryshevToth2017PRD96}
\begin{equation}
  \mu(\rho,\lambda)=\mu_0 J_0^2\!\Big(k\rho\sqrt\frac{2r_g}{z}\Big),
  \qquad
  \mu_0\simeq \frac{4\pi^2r_g}{\lambda},
  \label{eq:mu}
\end{equation}
where $k=2\pi/\lambda$ and $\rho$ is transverse separation in the SGL image plane \cite{Turyshev2017PRD95,TuryshevToth2017PRD96}.  The first zero of the monochromatic monopole response occurs at
\begin{equation}
  \rho_1=\frac{2.4048}{k}\sqrt{\frac{z}{2r_g}}
  =4.91\,\mathrm{cm}\Big(\frac{\lambda}{1\,\um}\Big)
   \Big(\frac{z}{650\,\AU}\Big)^{1/2}.
  \label{eq:first_zero}
\end{equation}
A metre-class telescope therefore averages over many oscillations at optical wavelengths.  For a circular collecting aperture of diameter $d$, aperture averaging over the Bessel oscillations gives \cite{TuryshevToth2020Photometric,TothTuryshev2021Recovery}
\begin{equation}
  \bar\mu_d(d,\lambda,z)=\mu_0\Big(J_0^2(x)+J_1^2(x)\Big),\qquad
  x=\frac{kd}{2}\sqrt{\frac{2r_g}{z}},
  \label{eq:aperture_average_gain}
\end{equation}
which approaches
\begin{equation}
  \bar\mu_d\rightarrow \frac{4b}{d}
  =3.03\times10^9\Big(\frac{1\,\m}{d}\Big)
   \left(\frac{z}{650\,\AU}\right)^{1/2}
\end{equation}
for $d\gg\rho_1$.  This finite-aperture gain, not the point-source peak alone, is the relevant local-telescope gain scale used in Eq.~(\ref{eq:geff}).  The characteristic formal angular response is
\begin{equation}
  \theta_{\rm SGL}\simeq0.38\frac{\lambda}{b}
  =0.103\Big(\frac{\lambda}{1\,\um}\Big)
   \Big(\frac{650\,\AU}{z}\Big)^{1/2}\nas.
  \label{eq:theta_sgl}
\end{equation}
Using $1.22\lambda/D_{\rm eq}=\theta_{\rm SGL}$, the equivalent filled-aperture diffraction diameter is
\begin{equation}
  D_{\rm eq}\simeq \frac{1.22\lambda}{\theta_{\rm SGL}}\simeq3.2b
  \simeq2.4\times10^6\Big(\frac{z}{650\,\AU}\Big)^{1/2}\km.
  \label{eq:deq}
\end{equation}
This equivalent baseline is millions of kilometres at optical wavelengths. It is not a physical collecting aperture and should not be used to infer a conventional field of view (FOV), but it correctly indicates the angular-resolution regime that becomes accessible if the SGL response can be calibrated.

At $1.3\,\mathrm{mm}$, Eq.~(\ref{eq:theta_sgl}) gives $\theta_{\rm SGL}\simeq0.13\,\uas$ at $650\,\AU$.  This is still far finer than the $\sim20\,\uas$ imaging scale of the first Event Horizon Telescope (EHT) images \cite{EHT2019I,EHT2019IV,EHT2022SgrA}.  Long-wavelength SGL operation is therefore scientifically attractive for black-hole and AGN science, but it is a different instrument problem: solar thermal emission, coronal plasma, receiver noise, annular coupling, occultation, and calibration must be modeled at millimeter/sub-millimeter wavelengths rather than assumed from the optical case.

\subsection{Wavelength regimes and instrument interpretation}

Eq.~(\ref{eq:theta_sgl}) may be evaluated from sub-micron optical wavelengths through the infrared, submillimetre, and millimetre regimes.  The formal angular response scales linearly with wavelength and remains far below the resolution of conventional facilities over a broad spectral interval.  This does not mean that a single SGL instrument operates identically at all wavelengths.  The wavelength sets the solar foreground, detector technology, ring-extraction method, aperture averaging, achievable spectral resolution, and the importance of solar plasma.

The optical and near-infrared range, roughly $0.3$--$5\,\um$, is the most mature regime for stellar surfaces, white dwarfs, reflected-light exoplanets, AGN continuum structure, compact lensed sources, and many spatially resolved spectra \cite{CHARA2005,Monnier2007,Roettenbacher2016,TuryshevToth2022Spectral}.  The dominant limitations are solar-corona brightness, solar-limb suppression, detector calibration, and the wave-optical PSF.  The thermal-infrared range, roughly $5$--$30\,\um$, is attractive for warm dust, evolved-star winds, planet-forming subfields, thermal exoplanet emission, and AGN tori, but it introduces solar-thermal background, cryogenic detector, and occulter design requirements.  Far-infrared and sub-millimeter operation is more speculative but may access cold dust, molecular lines, compact star-forming knots, and selected high-brightness line sources.  Millimeter operation is scientifically compelling for black-hole and AGN synchrotron emission \cite{EHT2019I,EHT2019IV,EHT2022SgrA}, but it is a separate instrument problem because solar plasma, receiver noise, scattering, and solar thermal emission must be treated explicitly.  Thus, broad SGL wavelength coverage should be understood as a family of SGL observatory modes sharing the same gravitational optical element, not as a single unchanged instrument.

\subsection{Aperture-averaged PSF, scalar benchmark, and vector measurement operator}

\begin{figure*}[t]
  \centering
  \includegraphics[width=0.86\textwidth]{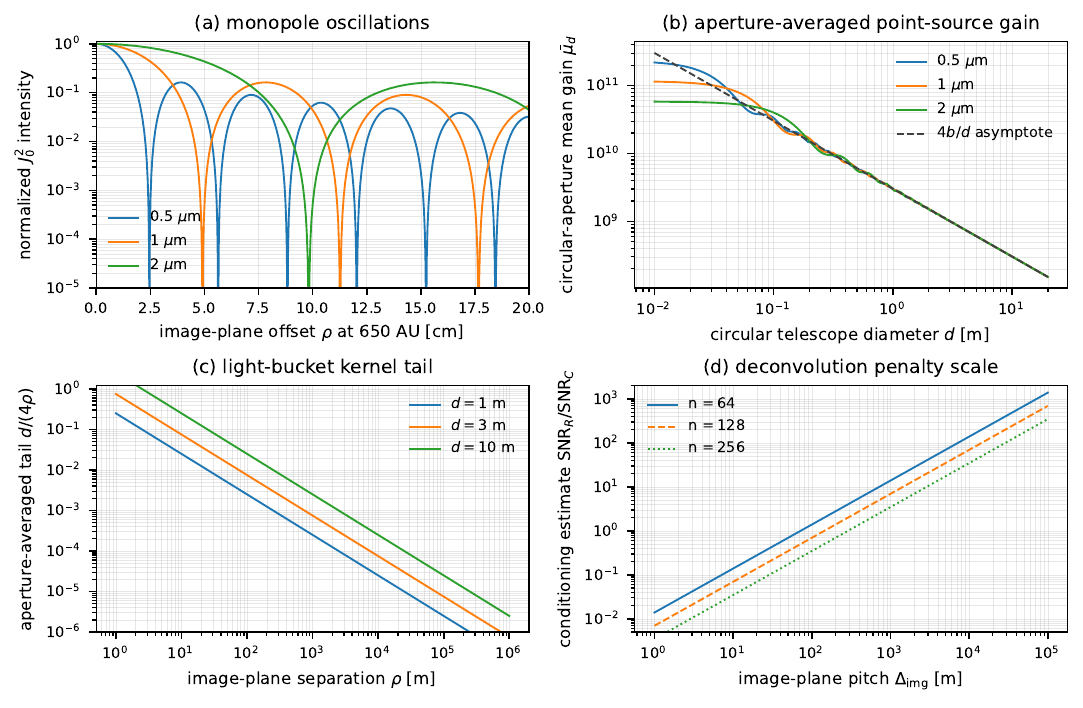}
  \caption{Monopole physical-optics and scalar-conditioning diagnostics. Panel (a) shows the normalized $J_0^2$ monopole response near the optical axis and the centimetre-scale first-zero structure. Panel (b) shows the exact circular-aperture mean gain $\bar\mu_d=\mu_0[J_0^2(x)+J_1^2(x)]$ and its $4b/d$ asymptote. Panel (c) shows the aperture-averaged light-bucket tail used in the scalar benchmark. Panel (d) gives the deconvolution penalty scaling of Eq.~(\ref{eq:snr_penalty}) as a function of image-plane pitch $\Delta_{\rm img}$ for linear raster dimensions $n=64$, $128$, and $256$. The figure separates physical-optics gain, aperture-integrated light-bucket imaging, and inverse-problem conditioning.}
  \label{fig:psf_conditioning}
\end{figure*}

The raw Bessel oscillations in Eq.~(\ref{eq:mu}) have centimeter-scale structure at optical wavelengths [see Fig~\ref{fig:psf_conditioning}(a)].  A meter-class telescope used as a photometric collector of the Einstein ring averages over many oscillations.  When the image-plane pitch is many aperture diameters, a useful aperture-averaged scalar light-bucket kernel is \cite{TuryshevToth2020Photometric,TothTuryshev2021Recovery}
\begin{equation}
  K(0)=1,\qquad K(\rho>0)\simeq\frac{d}{4\rho},
  \label{eq:kernel}
\end{equation}
with normalization set by the sampled field \cite{TuryshevToth2020Extended,TothTuryshev2021Recovery}.  This broad tail is the source of the SGL deconvolution problem.  The formal diffraction response may be sub-nanoarcsecond, but the delivered map resolution depends on the optical transfer function, sample pitch, measurement SNR, PSF knowledge, support constraints, and regularization.

The scalar convolution model used for the simulations is a controlled reduction of the general annular/vector SGL measurement operator developed for image recovery and spectral imaging \cite{TothTuryshev2021Recovery,TuryshevToth2022Spectral}.  A flight data stream should be modeled as
\begin{equation}
 y_{i\ell a p}\sim {\rm Poisson}\left[\int K_{i\ell a p}(\boldsymbol{\xi},t;\boldsymbol{\eta})
 I(\boldsymbol{\xi},\lambda_\ell,p,t)\,d^2\xi+b_{i\ell a p}(\boldsymbol{\eta})\right],
 \label{eq:vector_measurement}
\end{equation}
where $i$ indexes image-plane position, $\ell$ wavelength, $a$ ring azimuth or extraction sector, $p$ polarization, $I$ the source brightness field, $K$ the SGL+telescope+occulting response, $b$ the solar/coronal/instrumental/local background, and $\boldsymbol{\eta}$ nuisance parameters including spacecraft state, target ephemeris, solar multipoles, plasma, detector calibration, and ring-extraction geometry.  The scalar simulations collapse $(\ell,a,p,t)$ to one effective channel and replace Eq.~(\ref{eq:vector_measurement}) by a stationary convolution.  This is appropriate for comparing target classes and inverse-problem conditioning, but it is not a mission-grade vector, spectropolarimetric, time-dependent forward model. Consistent with the resolved exoplanet-imaging benchmark \cite{Turyshev2026DirectHighResExo}, the scalar annular statistic used here assumes that the focal-plane ring-extraction problem has already been reduced to an effective throughput $\eta_{\rm ann}$ and calibrated background statistic.  It does not demonstrate internal-coronagraph, external-occulter, ring-sector, diffraction-propagation, coronal-subtraction, or leakage-covariance performance; those terms enter the mission-level operator $K_{i\ell ap}$ and nuisance vector $\boldsymbol{\eta}$ in Eq.~(\ref{eq:vector_measurement}).

The scalar model is therefore not an alternative physical theory of the SGL measurement; it is the reduced statistic that a calibrated annular/vector instrument must deliver for a specified target, wavelength band, and observing mode.  The technology question is whether marginalizing or fitting the nuisance vector $\eta$ in Eq.~(\ref{eq:vector_measurement}) yields the same science observables---morphology, contrast, $\FRC_{50}$, flux normalization, and spectropolarimetric parameters---as the scalar benchmark to within the target error budget.  In this sense, the scalar simulations below define the information-recovery target for the full Poisson operator: they show which source classes are worth carrying into annular/vector propagation, not which flight data product has already been validated.

A statistically explicit flight estimator should begin with the Poisson likelihood rather than with a Gaussian least-squares approximation.  In a compact notation with source vector $\bm{x}$, measurement vector $\bm{y}$, response matrix $A(\bm{\eta})$, and background $\bm{b}(\bm{\eta})$, one convenient form is
\begin{equation}
  \widehat{\bm{x}}=\arg\min_{\bm{x}\ge0}\bigg\{
  2\sum_j\left[(A\bm{x}+\bm{b})_j-y_j\ln(A\bm{x}+\bm{b})_j\right]
  +\alpha\|L\bm{x}\|_2^2+\beta\Phi(\bm{x})\bigg\},
  \label{eq:poisson_estimator}
\end{equation}
where $L$ may impose smoothness or temporal coherence and $\Phi$ can encode positivity, sparsity, support, low-rank dynamics, or a physical source model.  The Gaussian dynamic inverse problem used later in Eq.~(\ref{eq:joint_inverse}) is the high-count limit of this more general formulation.

For stationary scalar reconstructions, a transparent baseline is the Fourier/Wiener inverse
\begin{equation}
  \widehat{O}(\bm{f})=\frac{\widehat{K}^{*}(\bm{f})}
  {|\widehat{K}(\bm{f})|^2+\gamma}\widehat{y}(\bm{f}),
  \label{eq:wiener}
\end{equation}
with support, non-negativity, and photometric constraints.  A useful conditioning estimate is

\begin{equation}
  \frac{\SNRr}{\SNRc}\simeq0.891\frac{\Delta_{\rm img}}{dn},
  \label{eq:snr_penalty}
\end{equation}
where $\SNRc$ is the convolved-raster signal-to-noise ratio, $\SNRr$ is a reconstructed-map residual SNR statistic, $\Delta_{\rm img}$ is the image-plane pitch, $d$ is telescope diameter, and $n$ is the linear raster dimension used in the resolved exoplanet-imaging benchmark \cite{Turyshev2026DirectHighResExo,TothTuryshev2021Recovery,TuryshevToth2022Resolved}.  The total number of image-plane samples is $N=n^2$.  Eq.~(\ref{eq:snr_penalty}) is not a universal reconstruction theorem; it is a practical warning that \emph{unconstrained} deconvolution can amplify noise and that delivered resolution is not equal to the formal SGL diffraction response.  Because it describes a regularized inverse with no support, positivity, or photometric constraint, it is \emph{not} expected to reproduce the masked, constrained residual statistic $\SNRr$ tabulated in Table~\ref{tab:metrics} (Eq.~\ref{eq:snrr_def}); evaluated on the benchmark parameters it reproduces the tabulated $\SNRr/\SNRc$ to within a factor $\sim\!1.3$ only for the white-dwarf case ($0.22$ predicted versus $0.28$ tabulated) and differs by factors of $\sim\!5$--$15$ for the others, including the M87*-like case where the constrained inverse yields $\SNRr>\SNRc$. These differences reflect the distinct definitions rather than an inconsistency.

A real solar lens is not a monopole.  Solar oblateness, rotation, and zonal harmonics perturb the gravitational phase, and extended-Sun calculations show caustics and ring distortions for compact sources \cite{TuryshevToth2021ExtendedSun,TuryshevToth2023Faint,Pijpers1998,Mecheri2004}.  Production SGL astronomy therefore requires a wavelength-dependent PSF library,
\begin{equation}
  K(\rho,\phi;\lambda,z,J_2,J_4,\ldots,\hat{\bm{s}}_\odot,\bm{e}_{\rm targ}),
  \label{eq:psf_library}
\end{equation}
where $\hat{\bm{s}}_\odot$ is the solar spin geometry and $\bm{e}_{\rm targ}$ represents target ephemeris and line-of-sight state.  For compact non-exoplanet targets, PSF model error is likely to be as important as photon noise. The few-percent anisotropic mismatch fields used in the scalar benchmarks, and the $J_2/J_4$-anchored white-dwarf closure test below, are therefore residual-sensitivity tests of the inverse problem.  They are not substitutes for the physical wave-optical solar-multipole, plasma, extended-Sun, and focal-plane ring-response library represented schematically by Eq.~(\ref{eq:psf_library}).

\subsection{Radiometry, spectroscopy, and polarimetry}

A compact-source photon-rate estimate remains useful if finite-source and aperture/ring-throughput limits are applied explicitly, as in resolved SGL exoplanet and spectral-imaging studies \cite{TuryshevToth2022Resolved,Turyshev2026DirectHighResExo,TuryshevToth2022Spectral,Turyshev2025DirectHighRes}.  For a continuum source with AB magnitude $m_{\rm AB}$, telescope diameter $d$, throughput $\eta$, and fractional bandwidth $\beta=\Delta\nu/\nu$, the unlensed photon rate is approximately
\begin{equation}
  Q_{\rm tel}=4.30\times10^{10}\,\eta\beta
  \Big(\frac{d}{1\,\m}\Big)^2 10^{-0.4m_{\rm AB}}\;\s^{-1}.
  \label{eq:q_ab}
\end{equation}
The SGL-coupled annular source rate is therefore
\begin{equation}
  Q_{s,\ell}=Q_{{\rm tel},\ell}\,
  \eta_{{\rm ann},\ell}
  \min\Big[\bar\mu_{d,\ell},\frac{4b}{D_{{\rm img},\ell}},\mu_{0,\ell}\Big].
  \label{eq:qs_eff}
\end{equation}
Eq.~(\ref{eq:qs_eff}) is the revised form of the earlier compact expression $Q_s={\cal G}_{\rm eff}Q_{\rm tel}$.  The compact expression is valid only when ${\cal G}_{\rm eff}$ has already been computed for the target, wavelength, aperture, selected field, and annular extraction strategy.  A point-like source, a $4\,\km$ white dwarf, a $438\,\km$ stellar disk, and a $3.4\times10^5\,\km$ full protoplanetary disk do not share the same effective gain.  

The convolved-raster SNR for dwell time $t$ is
\begin{equation}
  \SNRc=\frac{Q_st}{\sqrt{(Q_s+Q_b)t+\sigma_{\rm det}^2+(\epsilon_{\rm cal}Q_bt)^2+(\epsilon_{\rm sys}Q_st)^2}},
  \label{eq:snrc_general}
\end{equation}
where $Q_b$ contains solar corona, solar thermal leakage, zodiacal light, detector backgrounds, host-system leakage, and astrophysical foregrounds.  Bright stellar and white-dwarf cases are generally calibration-, PSF-, and variability-limited rather than photon-limited.  Faint planet-forming subfields are feasible only when the selected subfield is bright, the telescope aperture and dwell are adequate, and solar/background suppression is strong.

The \emph{photon-limited} convolved-raster SNR in a dwell $t$---i.e.\ Eq.~(\ref{eq:snrc_general}) with $\sigma_{\rm det}=\epsilon_{\rm cal}=\epsilon_{\rm sys}=0$---is
\begin{equation}
  \SNRc^{\rm phot}(t)=\frac{Q_s t}{\sqrt{(Q_s+Q_b)t}},
  \qquad
  t_{30}=30^2\frac{Q_s+Q_b}{Q_s^2},
 \label{eq:snrc_phot}
\end{equation}
where $t_{30}$ is the dwell at which $\SNRc^{\rm phot}=30$. We distinguish $\SNRc^{\rm phot}$ from the \emph{effective} information floor $\SNRc^{\rm eff}$ tabulated in Table~\ref{tab:sim_assumptions} and used for the reconstructions: the latter includes the calibration, systematic, and PSF-mismatch terms of Eq.~(\ref{eq:snrc_general}) and is, for the bright cases, far below $\SNRc^{\rm phot}$ (e.g.\ for the white dwarf $\SNRc^{\rm phot}\sim10^4$--$10^5$ per sample versus the adopted $\SNRc^{\rm eff}=95$). The reconstruction fidelity is therefore set by systematics, not by the photon rates in Table~\ref{tab:self_luminous_rates}. For the reflected exo-Earth benchmark, $\SNRc^{\rm phot}(1\,\s)\simeq1.0$ and $t_{30}\simeq8.7\times10^2\,\s$.

Spatially resolved spectroscopy and polarimetry are natural SGL products, but each spectral or polarization channel has its own PSF, gain, background, detector response, and source morphology.  For continuum channels, photon rates scale approximately as $R_\lambda^{-1}$ at fixed resolving power $R_\lambda=\lambda/\Delta\lambda$.  Spectropolarimetric products such as stellar velocity fields, Zeeman maps, AGN line profiles, disk kinematics, and compact narrowband line searches \cite{DonatiLandstreet2009,WickramasingheFerrario2000,Ferrario2015,Gravity2018BLR,Andrews2018} are therefore coupled inference problems rather than independent narrow-band images.

\section{Target phase space and observability metrics}
\label{sec:phase_space}

The SGL target phase space is governed by five quantities: angular size, image-plane size, flux/background ratio, variability time, and uniqueness relative to other facilities.  We use these to classify candidate science cases before detailed mission design. A compact way to state the closure test for a candidate target is
\begin{equation}
  {\cal C}=\left(D_{\rm img},\;\mu_{\rm ext},\;Q_s/Q_b,\;\chi_t,\;\sigma_\rho/\Delta_{\rm img},\;\epsilon_{\rm cal}Q_b/Q_s,\;\delta K/K,\;U\right),
  \label{eq:closure_vector}
\end{equation}
where $\delta K/K$ denotes residual PSF error after calibration and $U$ is the uniqueness ratio defined below. A target class is high priority only if the components of ${\cal C}$ close simultaneously: it must be compact enough to scan, bright enough to calibrate, slowly varying enough to model, recoverable under plausible PSF and metrology residuals, and scientifically unique relative to other facilities.

\subsection{Image-plane extent}

The first discriminator is $D_{\rm img}=z\Theta$.  The following regimes are useful at $650\,\AU$:
\begin{center}
\begin{tabular}{ll}
\toprule
Image-plane diameter & Observing interpretation \\
\midrule
$<1\,\m$ & effectively point-like; better for amplified photometry, astrometry, or spectroscopy \\
$1\,\m$--$10\,\km$ & compact imaging regime; excellent for white dwarfs and compact lensed sources \\
$10$--$10^3\,\km$ & long-scan imaging; plausible for stars, black-hole rings, AGN BLR, subfields \\
$10^3$--$10^5\,\km$ & sparse/subfield regime; full imaging requires many spacecraft or long campaigns \\
$>10^5\,\km$ & generally impractical for full-field imaging \\
\bottomrule
\end{tabular}
\end{center}
This classification is not an absolute field-of-view limit.  A sufficiently capable SGL mission could scan an extended target across the image plane and reconstruct a resolved image or spatially resolved spectrum.  The issue is cost and conditioning: scan time grows with the number of samples, finite-source gain decreases as the selected field grows, and target variability or background structure becomes harder to model.  The early high-return domain therefore favors compact stellar remnants, ordinary stellar disks, black-hole environments, AGN inner structures, small planet-forming subfields, and selected compact components of otherwise extended systems.

\paragraph*{Amplified non-imaging mode.}
Not every scientifically valuable SGL target must be spatially resolved.  If $D_{\rm img}\lesssim d$ or $D_{\rm img}\lesssim\Delta_{\rm img}$, raster imaging is poorly conditioned or unnecessary, but the target can still exploit the aperture-averaged SGL gain as an amplified photometric, spectroscopic, polarimetric, or timing source.  This ``point-source'' mode is relevant for nearby neutron stars, quiescent black-hole
X-ray binaries, compact broad-line regions, and narrow-line emitters: the science observable is then a high-SNR spectrum, pulse profile, reverberation signal, radial velocity, polarization curve, or line centroid rather than a reconstructed image
\cite{Walter1996,OzelFreire2016,RemillardMcClintock2006,CasaresJonker2014}.  Such targets should not be promoted as resolved-imaging cases, but they may be valuable
secondary or piggyback targets on a preselected focal line.

\subsection{Variability and dynamic inversion}

The second discriminator is temporal coherence.  A single-spacecraft raster is assembled sequentially, so an object with variability time $t_{\rm var}$ is quasi-static only if
\begin{equation}
  \chi_t=\frac{T_{\rm scan}}{t_{\rm var}}\ll1.
  \label{eq:chi}
\end{equation}
When $\chi_t\sim1$ or larger, the reconstruction must be dynamic, as already recognized in rotating and time-variable SGL imaging problems \cite{TothTuryshev2023Rotating}.  This is not a minor correction for stellar granulation, flares, Sgr A*, AGN reverberation, protoplanetary accretion shocks, and transients.  Multi-spacecraft sampling reduces $T_{\rm scan}$ and provides independent phase coverage.  A general SGL observatory is therefore more plausibly a small formation or string of spacecraft than a single photometric probe, especially for time-domain and compact-object science \cite{Friedman2024,Helvajian2023}. The resolved exoplanet-imaging benchmark treats rotation and cloud fields as explicit controlled stress tests; here the same issue is represented by the generic temporal-coherence parameter $\chi_t$, with target-specific dynamic inversions deferred to mission-level studies for stars, disks, AGN, compact remnants, and transients.

\subsection{Uniqueness relative to other observatories}

The third discriminator is uniqueness.  Coronagraphs and starshades suppress glare but do not improve diffraction; space telescopes and extremely large telescopes remain limited by filled-aperture diameters; optical/IR interferometers provide sub-milliarcsecond resolution for bright targets but not nanoarcsecond mapping of compact surfaces; ALMA and ngVLA provide exquisite millimetre interferometry but not optical/NIR stellar-surface or white-dwarf maps; EHT/ngEHT provide tens of microarcsecond horizon-scale imaging, not sub-microarcsecond imaging of ring substructure \cite{CHARA2005,ALMAPartnership2015,Andrews2018,EHT2019I,EHT2019IV,EHT2022SgrA}.  We define a practical uniqueness ratio
\begin{equation}
  U=\frac{\theta_{\rm best,foreseeable}}{\theta_{\rm SGL,delivered}},
  \label{eq:unique}
\end{equation}
where $\theta_{\rm SGL,delivered}$ is not the formal diffraction limit but the scale recovered after sampling, background subtraction, and inversion.  Operationally we identify $\theta_{\rm SGL,delivered}$ with the truth-referenced $\FRC_{50}$ scale of Table~\ref{tab:metrics}; converted to angular units at the benchmark distances this corresponds to truth-referenced $\FRC_{50}$ scales (\emph{not} achieved on-orbit resolutions) of approximately $11\,\uas$ (solar analog, $1.7\times10^4\,\km$ at $10\,\pc$), $0.21\,\uas$ (white dwarf, $315\,\km$ at $10\,\pc$), $0.66\,\uas$ (M87*-like, reported directly), and $11\,\uas$ (protoplanetary subfield, $1.6\times10^{-3}\,\AU$ at $140\,\pc$).  We caution that $\FRC_{50}$ here is a benchmark diagnostic computed against a known truth, not a flight half-data resolution.  Relative to the best foreseeable comparators in Table~\ref{tab:facility_comparison}---sub-milliarcsecond optical interferometry for stellar surfaces and white dwarfs, and tens-of-microarcsecond VLBI for compact millimetre sources---these delivered scales correspond to $U\gg1$; we state the comparison qualitatively rather than as a single numeral because a defensible $\theta_{\rm best,foreseeable}$ is target- and epoch-dependent.

\begin{figure*}[t]
  \centering
  \includegraphics[width=0.86\textwidth]{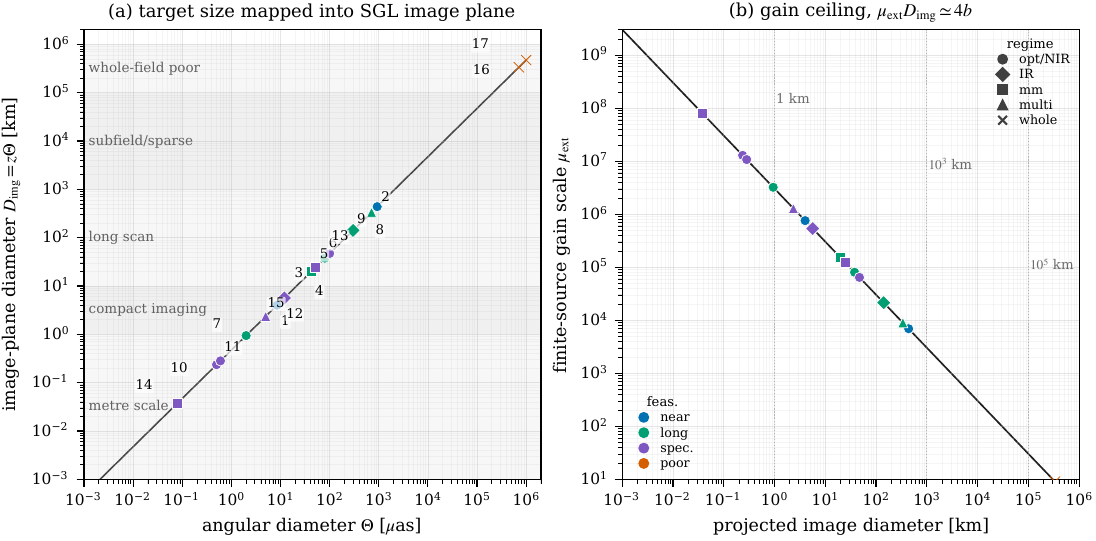}
  \caption{Target phase space and finite-source gain for representative SGL applications.  Panel (a) maps angular diameter into image-plane diameter at $650\,\AU$ and places the example IDs listed in Table~\ref{tab:target_examples}.  Panel (b) shows the finite-source gain ceiling $\mu_{\rm ext}D_{\rm img}\simeq4b$ using the same marker coding but omits repeated numeric IDs to preserve readability.  The high-$z$ marker denotes compact components such as nuclear star-forming knots, lensed stars, compact AGN, or LRD nuclei, not entire high-redshift galaxies. Color gives qualitative feasibility, and marker shape gives the dominant wavelength or observing regime.   The plotted examples are representative stress points in the observability space, not an exhaustive taxonomy of astrophysics or cosmology.}
  \label{fig:phase_space}
\end{figure*}

\begin{table*}[t]
\caption{Marker IDs used in Fig.~\ref{fig:phase_space}.  The table is a guide to the plotted examples; feasibility and wavelength regime are encoded by marker color and shape in the figure. The high-$z$ marker denotes compact components such as nuclear star-forming knots, lensed stars, compact AGN, or LRD nuclei, not entire high-redshift galaxies.}
\label{tab:target_examples}
\centering
\renewcommand{\arraystretch}{1.08}
\begin{tabular}{c l l l c l l l}
\toprule
ID & Example & Feas. & Regime & ID & Example & Feas. & Regime \\
\midrule
1 & white dwarf, $10\,\pc$ & near & optical/NIR & 10 & compact lensed knot/star & spec. & optical/NIR \\
2 & solar analog, $10\,\pc$ & near & optical/NIR & 11 & Cepheid in M31 & spec. & optical/NIR \\
3 & solar twin, $100\,\pc$ & long & optical/NIR & 12 & red supergiant in M31 & spec. & IR \\
4 & M87*-scale shadow & long & sub-mm/mm & 13 & high-$z$ compact source/LRD nucleus & spec. & optical/NIR--IR \\
5 & Sgr A* & spec. & sub-mm/mm & 14 & compact maser/line knot & spec. & sub-mm/mm \\
6 & AGN broad-line region & long & optical/NIR & 15 & predictable transient photosphere & spec. & multi-band \\
7 & compact AGN disk & long & optical/UV & 16 & full $100\,\AU$ disk & poor & whole field \\
8 & $0.1\,\AU$ disk subfield & long & IR/multi-band & 17 & full $1''$ lens system & poor & whole field \\
9 & inner wind/dust zone & long & IR & & & & \\
\bottomrule
\end{tabular}
\renewcommand{\arraystretch}{1.0}
\end{table*}

\begin{figure*}[t]
  \centering
  \includegraphics[width=0.86\textwidth]{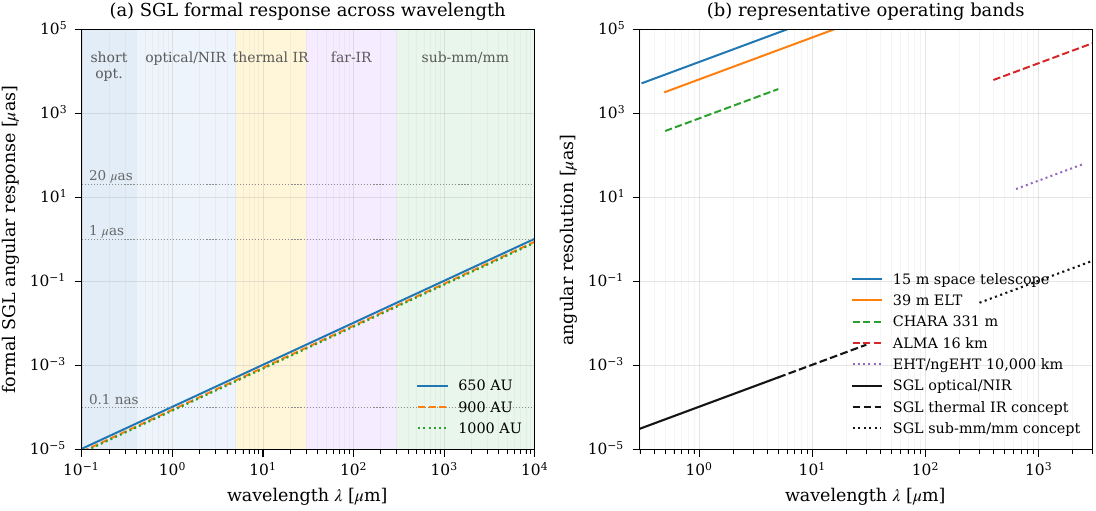}
  \caption{SGL angular-response and facility-comparison scalings over wavelength.  Panel (a) shows the formal SGL response from short optical wavelengths through infrared, far-infrared, submillimetre, and millimetre regimes for $650$, $900$, and $1000\,\AU$.  The shaded bands identify approximate wavelength regimes; they are not equal-maturity instrument bands.  Panel (b) compares SGL concept regimes with representative foreseeable facilities over their approximate operating ranges.  The SGL curve may be evaluated broadly, but practical performance at each wavelength requires a wavelength-specific treatment of solar corona, solar thermal background, plasma, detectors, occultation, annular extraction, and calibration.}
  \label{fig:facility_resolution}
\end{figure*}

\section{Simulation methodology and benchmark assumptions}
\label{sec:simulation_method}

The simulations  in this paper isolate SGL observability and scalar inverse conditioning under traceable assumptions.  Each case is mapped into the SGL image plane, propagated through an aperture-averaged scalar kernel, corrupted by stochastic noise and structured residuals, reconstructed with a non-oracle Wiener/Tikhonov inverse, and evaluated with common masks and metrics.  The validation checks are positivity and support consistency, masked residuals, support leakage, correlation coefficient, contrast recovery, radial spatial-power recovery, $\FRC_{50}$, and transfer/conditioning diagnostics.  Together these checks determine whether the selected astrophysical morphology survives the scalar SGL convolution and deconvolution with useful spatial information.

The benchmark assumptions are intentionally explicit.  The source is static during a raster; the channel is an effective scalar bandpass; the SGL response is stationary over the sampled field; the image-plane coordinates are known; the support mask is an astrophysical prior rather than a fitted parameter; and the injected $\SNRc$ is an effective information quality after photon, background, calibration, PSF, and variability floors.  These assumptions define a controlled recovery problem.  The corresponding observatory problem is obtained by replacing the scalar operator with Eq.~(\ref{eq:vector_measurement}) and fitting or marginalizing the nuisance parameters in Eq.~(\ref{eq:poisson_estimator}).

\subsection{Common scalar forward model}

For each source map $O$, normalized as specified in Appendix~\ref{app:benchmark_details}, we compute
\begin{equation}
  y=H_{\rm true}O+b+n,
  \label{eq:sim_forward}
\end{equation}
where $H_{\rm true}$ is the aperture-averaged scalar SGL convolution operator. The inverse assumes a calibrated monopole operator $H_0$, so the benchmark includes an unmodeled low-order PSF perturbation
\begin{equation}
  K_{\rm true}(\rho,\phi)=K_0(\rho)
  \Big(1+\epsilon_2\cos2(\phi-\phi_0)+\epsilon_4\cos4(\phi-\phi_0)\Big),
  \label{eq:sim_psf_mismatch}
\end{equation}
renormalized to conserve total kernel weight. This is not a solar-multipole calculation; it is a controlled PSF-mismatch sensitivity test that probes energy-conserving anisotropic power in the aperture-averaged scalar tail. The values used are $\epsilon_2=0.012$, $0.018$, $0.040$, and $0.030$ for the stellar, white-dwarf, M87*, and protoplanetary-subfield simulations, with $\epsilon_4=0.35\epsilon_2$. Because $\epsilon_2$ is chosen per scene, the four headline SSIM values are not directly comparable as relative feasibility: the M87*-like case carries the largest mismatch ($\epsilon_2=0.040$) yet the second-highest SSIM, because its high-contrast crescent dominates the metric. A target-specific study should add at least one reconstruction per scene at a common $\epsilon_2$ to separate intrinsic conditioning from scene contrast. The benchmark SSIM values should therefore be interpreted as controlled-mismatch reconstruction metrics, not as validated performance predictions for a realistic solar PSF. Physical validation requires the wave-optical solar-multipole and plasma PSF library in Eq.~(\ref{eq:psf_library}).

As a targeted bridge between this generic stress test and a physical solar PSF, we also define a low-order solar-multipole residual surrogate for the white-dwarf grid.  This surrogate is not a wave-optical propagation of an oblate, rotating, plasma-filled Sun; it is a scalar kernel displacement anchored to the expected image-plane scale of the zonal harmonics.  The scale follows by multiplying a zonal-harmonic bending perturbation of order $\delta\alpha_n\sim 2n r_g J_n R_\odot^n/b^{n+1}$ by the local focal distance $z=b^2/(2r_g)$, giving an image-plane displacement scale $a_n\sim n J_n R_\odot^n/b^{n-1}$.  This motivates the quadrupole and hexadecapole amplitudes used below without claiming that the true PSF is a translated monopole kernel.  We write
\begin{equation}
  K_J(\rho,\phi;f_J)=K_0\big(|\rho-\delta\rho_J(\phi)|\big),
  \label{eq:j_proxy_kernel}
\end{equation}
with
\begin{equation}
  \delta\rho_J(\phi)=f_J\Big(a_2\cos2(\phi-\phi_\odot)+a_4\cos4(\phi-\phi_\odot)\Big),\qquad
  a_2=2|J_2|\frac{R_\odot^2}{b},\qquad
  a_4=4|J_4|\frac{R_\odot^4}{b^3}.
  \label{eq:j_proxy_scale}
\end{equation}
Here $f_J$ is the fraction of the deterministic low-order multipole displacement left unmodeled after calibration.  For the numerical closure test we adopt fiducial helioseismic values $J_2=2.2\times10^{-7}$ and $|J_4|=4.0\times10^{-9}$ \cite{Pijpers1998,Mecheri2004}; at $650\,\AU$ and $b=1.089R_\odot$, these give $a_2=2.8\times10^2\,\m$ and $a_4=8.6\,\m$.  Because $a_4/a_2\simeq0.03$, the closure test is quadrupole-dominated and essentially insensitive to the precise value of $J_4$, which is in any case far less well constrained observationally than $J_2$. The white-dwarf pitch is $\Delta_{\rm img}=31.4\,\m$, so the raw quadrupole-scale residual is many pixels across and cannot simply be ignored.  The coefficients $a_2$ and $a_4$ are used only to set the image-plane scale of a calibrated residual proportional to the low-order zonal-harmonic deflection scale.  Eq.~(\ref{eq:j_proxy_kernel}) should not be interpreted as a literal radial translation of the true solar multipole PSF, nor as a caustic calculation.  It asks a narrower and useful closure question: if deterministic low-order multipole structure is calibrated to a residual fraction $f_J$, is a scalar inverse stable when the remaining image-plane error is a specified fraction of the raster pitch?  The relevant flight question is therefore not whether the Sun has a nonzero quadrupole--it does--but what residual fraction remains after a solar-multipole PSF library and ring-calibration procedure have removed the deterministic component.

Noise is injected as a combination of Gaussian photon noise, low-frequency structured residuals, and systematic floors. The adopted $\SNRc$ values are not raw photon limits. They represent convolved-raster information quality after imposing photon statistics, source/background ratio, calibration residuals, unmodeled PSF structure, and source-variability floors according to Eq.~(\ref{eq:snrc_general}). This convention is necessary because the brightest non-exoplanet targets are photon-rich under plausible SGL gains; photon statistics alone would overstate the fidelity of the reconstruction.

The reconstruction uses Eq.~(\ref{eq:wiener}) with the nominal monopole aperture-averaged kernel. The regularization parameter is selected from the adopted noise level and kernel transfer function, not from the hidden truth. Disk-like cases are constrained to non-negative values within the source support; field-like cases are constrained to non-negative surface brightness over the selected subfield. The implementation details needed to reproduce the scalar benchmarks are collected in Appendix~\ref{app:benchmark_details} so that the present paper does not depend on an unpublished companion manuscript.

\subsection{Metric definitions and masks}
\label{subsec:metrics}

All image-domain metrics are computed on a specified metric mask $M$. For stellar and white-dwarf disks, $M$ is the projected disk support after applying the parity convention and excluding pixels outside the limb. For the M87*-like scene, $M$ is the union of the compact ring, jet-base support, and a small surrounding background annulus used to penalize false structure. For the protoplanetary subfield, $M$ is the full selected subfield unless otherwise stated. The same mask is used for NRMSE, correlation coefficient, contrast recovery, and $\SNRr$.

The normalized root-mean-square error is
\begin{equation}
  {\rm NRMSE}=\frac{\left[|M|^{-1}\sum_{i\in M}\left(O_{{\rm rec},i}-O_{{\rm true},i}\right)^2\right]^{1/2}}
  {O_{{\rm max,true}}-O_{{\rm min,true}}},
  \label{eq:nrmse_def}
\end{equation}
where the denominator is the true dynamic range on the same mask. SSIM~\cite{Wang2004SSIM} is computed on the same displayed dynamic range after applying the same photometric normalization to the reconstruction and truth. The contrast recovery is
\begin{equation}
  C_{\rm rec}/C_{\rm true}=\frac{P_{95}(O_{\rm rec}|M)-P_5(O_{\rm rec}|M)}{P_{95}(O_{\rm true}|M)-P_5(O_{\rm true}|M)},
  \label{eq:contrast_def}
\end{equation}
where $P_q$ denotes the $q$th percentile. $\FRC_{50}$ is the spatial scale at which the Fourier-ring correlation~\cite{vanHeelSchatz2005} between reconstruction and truth first falls below 0.5 after azimuthal binning in the appropriate source-plane units.

The convolved-raster SNR, $\SNRc$, is the effective measurement SNR of the scalar raster before deconvolution. The reconstructed-map SNR, $\SNRr$, is a post-reconstruction residual statistic over $M$,
\begin{equation}
  \SNRr=\frac{\langle O_{\rm true}\rangle_M}{\left[|M|^{-1}\sum_{i\in M}\left(O_{{\rm rec},i}-O_{{\rm true},i}\right)^2\right]^{1/2}},
  \label{eq:snrr_def}
\end{equation}
with the mean and residual evaluated after the same photometric normalization used for the image-domain metrics. It is not intended to be predicted directly by Eq.~(\ref{eq:snr_penalty}). Eq.~(\ref{eq:snr_penalty}) is a conditioning warning for unconstrained deconvolution; Eq.~(\ref{eq:snrr_def}) is a masked residual statistic after support, positivity, and photometric constraints. Cases in which $\SNRr>\SNRc$ are therefore not contradictions; they indicate that the reported residual statistic and the convolved-raster information metric measure different quantities.

Because support masks can otherwise hide false structure, we also report a support-leakage diagnostic for compact-mask cases,
\begin{equation}
  f_{\rm out}=\frac{\sum_{i\notin M}\max(O_{{\rm rec},i},0)}{\sum_i\max(O_{{\rm rec},i},0)+\epsilon_F},
  \label{eq:fout_def}
\end{equation}
where $\epsilon_F$ is a numerical guard much smaller than the reconstructed total positive flux.  For a whole selected field, such as the protoplanetary subfield benchmark, the metric support is the field itself and $f_{\rm out}$ is not a separate diagnostic.  For disk, ring, and compact-component cases, $f_{\rm out}$ measures how much positive reconstructed power appears outside the assumed science support.

\subsection{Benchmark scenes}

The four benchmark scenes are deliberately distinct:
\begin{enumerate}
\item A solar analog at $10\,\pc$ with limb darkening, granulation, starspots, plage-like bright regions, and a phenomenological surface velocity field.  The angular diameter is $0.93\,\mas$ and $D_{\rm img}=438\,\km$.

\item A magnetic white dwarf at $10\,\pc$ with an Earth-size diameter, limb darkening, hot/cool spots, a phenomenological accretion belt, and a smooth Zeeman/Stokes magnetic-field map.  The angular diameter is $8.5\,\uas$ and $D_{\rm img}=4.02\,\km$.

\item An M87*-like millimetre source with a crescent ring, hot spot, shadow depression, and jet-base extension.  The angular diameter is $42\,\uas$ and $D_{\rm img}=19.8\,\km$.

\item A $0.1\,\AU$ planet-forming subfield at $140\,\pc$ containing a phenomenological circumplanetary disk, spiral wake, gap-edge gradient, dust trap, and accretion-shock component.  The angular size is $0.714\,\mas$ and $D_{\rm img}=337\,\km$.
\end{enumerate}

These scenes are analytic, not decorative.  They are chosen because each maps to a specific science measurement: stellar activity and velocity structure, white-dwarf temperature/magnetic mapping, black-hole/jet morphology, and sub-AU planet-formation structure.  The benchmark set is intentionally minimal: it spans a large stellar disk, a compact optical/NIR remnant, a compact long-wavelength relativistic source, and an extended but selected planet-forming subfield.  Other target classes in Sec.~\ref{sec:science_cases} are assessed through the same phase-space and radiometric closure tests rather than by adding redundant gallery rows.  The scenes are deliberately analytic: replacing them with stellar-MHD, white-dwarf atmosphere and spectropolarimetric, GRMHD, BLR, and radiative-transfer scene libraries is a required step for target-specific mission validation.

\begin{figure*}[t]
  \centering
  \includegraphics[width=0.90\textwidth]{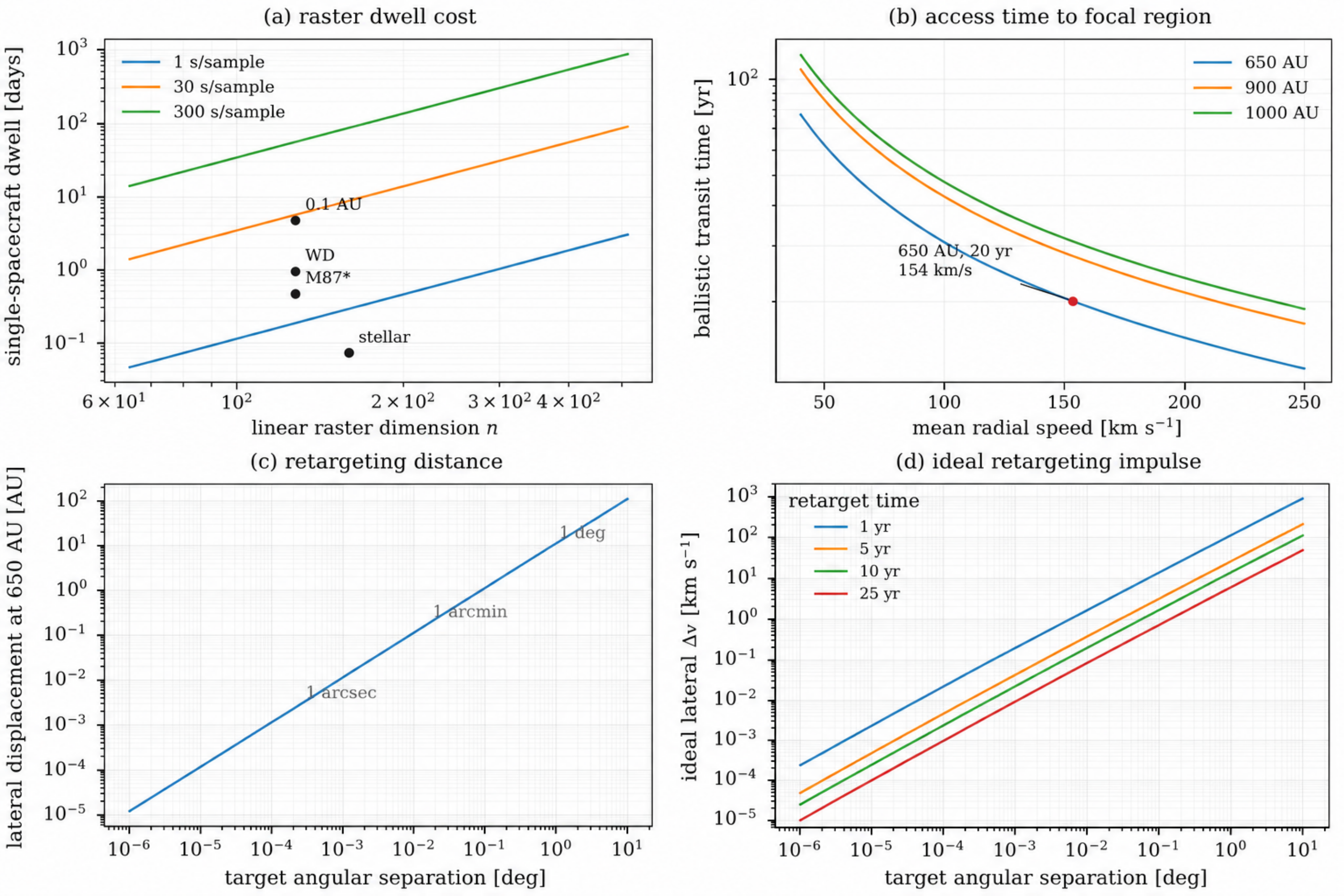}
  \caption{Mission implementation scalings.  Panel (a) shows cumulative dwell time versus linear raster dimension $n$ for representative sample dwells, before slews, repeats, and calibration. Panel (b) gives ballistic access time for $650$, $900$, and $1000\,\AU$ as a function of mean radial speed.  Panel (c) shows the lateral displacement required to retarget at $650\,\AU$.  Panel (d) gives an idealized triangular accelerate--decelerate lateral $\Delta v$ for selected retargeting times.  The SGL is operationally a focal-line observatory; rapid wide-angle retargeting requires advanced propulsion or multiple spacecraft.}
  \label{fig:mission_constraints}
\end{figure*}

Most scalar benchmarks use $n=128$ because it is a controlled power-of-two linear grid that resolves the source morphology while keeping the inverse problem transparent and the wall-dwell estimates comparable.  The solar analog uses $n=160$ because the projected disk is much larger and a coarser grid would undersample the imposed surface structure.  These choices are representative sampling points, not optimized mission designs; Eq.~(\ref{eq:snr_penalty}) and Fig.~\ref{fig:mission_constraints} show how conditioning and scan cost scale with $n$, $\Delta_{\rm img}$, and dwell time.

\subsection{Radiometric assumptions}

Table~\ref{tab:sim_assumptions} gives the signal and observing assumptions.  The optical/NIR examples use $\eta=0.25$ and $\beta=0.1$, but the source rates now apply a finite-source gain ceiling using Eq.~(\ref{eq:mu_ext_product}) rather than a single target-independent SGL gain.  The stellar and white-dwarf examples remain calibration-limited because they are intrinsically bright.  The protoplanetary-subfield example is treated as a bright selected subfield, $m_{\rm AB}=18$, observed with a $3\,\m$ telescope; fainter subfields would require longer dwell, larger aperture, better background rejection, or line/thermal observing modes.  The M87*-scale example asks only whether compact ring/jet morphology is recoverable if a future long-wavelength SGL instrument delivers a specified scalar information quality.  It adopts $d=3\,\m$, $\lambda=1.3\,\mathrm{mm}$, a $0.7\,\mathrm{Jy}$ compact flux density, $8\,\mathrm{GHz}$ bandwidth, ${\cal G}_{\rm eff}=10^4$, and a residual millimetre background of $3\times10^{10}\,\s^{-1}$.  These are demonstration inputs, not validated mm-SGL performance estimates; receiver temperature, solar thermal emission, plasma phase, occultation, annular coupling, polarization leakage, and dynamic sampling remain outside the scalar benchmark.

\subsection{Noise floors and robustness sensitivities}
\label{subsec:robustness}

\begin{figure*}[t]
  \centering
  \includegraphics[width=0.90\textwidth]{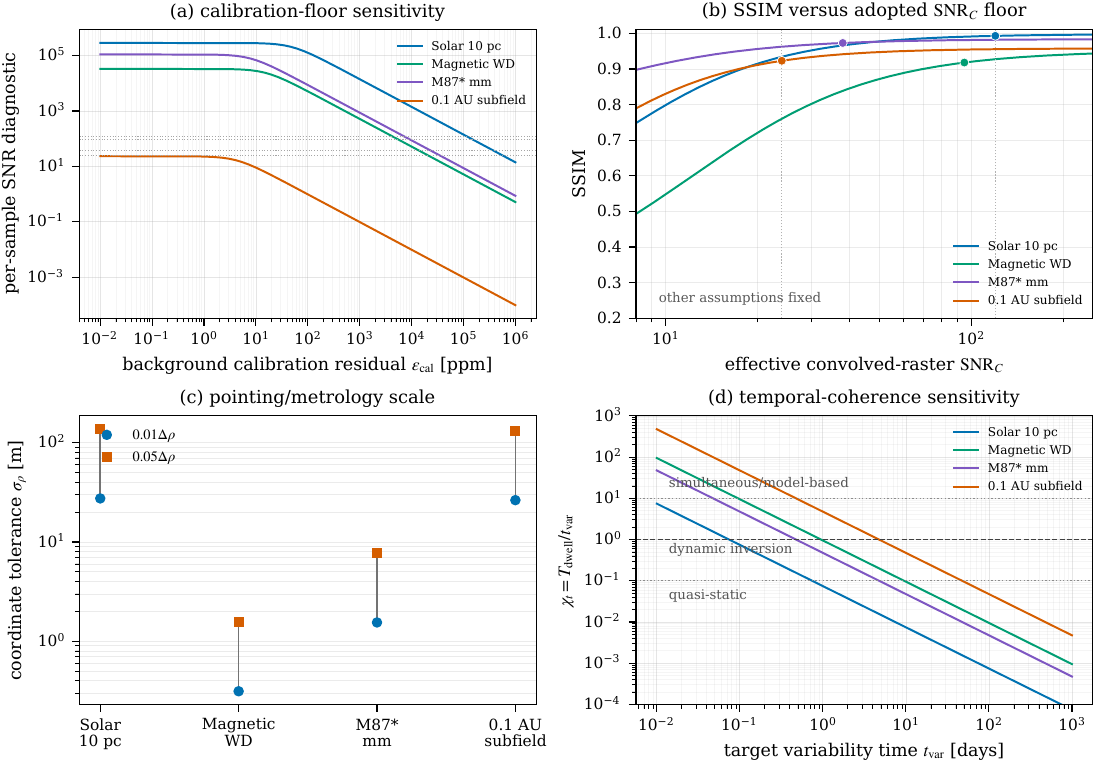}
  \caption{First-order sensitivity diagnostics for the four scalar benchmarks. Panel (a) uses Eq.~(\ref{eq:snrc_general}) to show the per-sample SNR degradation caused by a multiplicative background residual; dotted horizontal lines mark the adopted effective $\SNRc$ floors in Table~\ref{tab:sim_assumptions}. Panel (b) sweeps SSIM versus the imposed effective $\SNRc$ floor while holding the scalar scenes, kernels, support masks, and regularization convention fixed; filled markers identify the reported benchmark values. This panel isolates inverse-conditioning sensitivity to the assumed information floor and is not a derivation of $\SNRc$ from a detector/background model. Panel (c) shows coordinate-knowledge tolerances corresponding to $0.01\Delta_{\rm img}$ and $0.05\Delta_{\rm img}$. Panel (d) shows the dwell-only temporal-coherence ratio $\chi_t=T_{\rm dwell}/t_{\rm var}$ before slew, repeats, and calibration overhead. The figure maps robustness boundaries for the benchmark assumptions; it is not an end-to-end mission simulation.}
  \label{fig:sensitivity_boundaries}
\end{figure*}

The scalar benchmarks include explicit noise floors rather than assuming that photon statistics alone determine performance. The adopted $\SNRc$ values in Table~\ref{tab:sim_assumptions} should be read as effective information-quality levels after photon statistics, background subtraction, residual PSF mismatch, calibration drift, and source variability are combined. Alternative floor choices change the table entries primarily through Eq.~(\ref{eq:snrc_general}): in the background-dominated regime $\SNRc\propto Q_s\sqrt{t/Q_b}$ until a calibration or systematic term dominates, whereas in the calibration-limited regime additional photons do not improve the reconstruction unless $\epsilon_{\rm cal}$, $\epsilon_{\rm sys}$, or PSF mismatch are reduced.  To make the assumption dependence explicit, Fig.~\ref{fig:sensitivity_boundaries}(b) sweeps the effective $\SNRc$ floor while holding the scalar scenes, kernels, masks, support priors, and regularization convention fixed.  The sweep is not a derivation of $\SNRc$ from a physical detector/background model; it is a compact inverse-conditioning diagnostic showing how rapidly the headline SSIM values would change if the imposed information floor were lower or higher.

Table~\ref{tab:sensitivity_closure} summarizes the first-order sensitivity tests that bound the current benchmarks. The highest-leverage item is PSF mismatch: the generic $\epsilon_2,\epsilon_4$ runs test numerical stability, while the $J_2/J_4$ closure test in Sec.~\ref{sec:sim_results} converts the same issue into an image-plane residual scale tied to solar physics. The table is not a substitute for a full mission simulator, but it makes the success and failure variables explicit. It also identifies which metrics carry the most physical information: SSIM is useful for comparing controlled scalar reconstructions, FRC$_{50}$ and contrast recovery are tied to delivered spatial information, and $f_{\rm out}$ tests whether support constraints are hiding artifacts. For planning-level closure we use three practical thresholds: residual low-order PSF structure should be calibrated below the science-contrast or approximately half-pixel scale, background residuals should satisfy $\epsilon_{\rm cal}Q_b/Q_s$ below the allowed false-signal fraction, and compact-source reconstructions should keep $f_{\rm out}\lesssim0.05$ unless the out-of-support flux is explicitly modeled as astrophysical background.

\begin{table*}[t]
\caption{First-order robustness and closure requirements for the scalar benchmarks. The entries define the minimum diagnostic sweeps needed before controlled scalar reconstructions are interpreted as mission predictions.}
\label{tab:sensitivity_closure}
\centering
\renewcommand{\arraystretch}{1.08}
\setlength{\tabcolsep}{4pt}
\begin{tabular*}{\textwidth}{@{\extracolsep{\fill}}llll}
\toprule
Variable & Current test & Failure scale & Diagnostic \\
\midrule
PSF mismatch & $\epsilon_2,\epsilon_4$ stress; WD $J_2/J_4$ residual & $\sim0.5\Delta_{\rm img}$ or science contrast & metrics versus residual \\
Support leakage & masked residuals; $f_{\rm out}$ & false flux comparable to signal & $f_{\rm out}$; morphology \\
Calibration/background & $\SNRc$ floor; residual $Q_b$ & $\epsilon_{\rm cal}Q_b/Q_s$ too large & SSIM versus $\SNRc$; required $\epsilon_{\rm cal}$ \\
Finite-source gain & Eq.~(\ref{eq:geff}) planning model & selected field too large & dwell and FRC$_{50}$ versus field \\
Metrology/pointing & coordinates assumed known & $\sigma_\rho\gtrsim0.01$--$0.05\Delta_{\rm img}$ & jitter, drift, smear sweeps \\
Raster pitch/dwell & one $n$, $\Delta_{\rm img}$, dwell per case & OTF cutoff or insufficient $\SNRc$ & $\FRC_{50}$, aliasing, contrast \\
Temporal variability & static scalar scene & $\chi_t=T_{\rm wall}/t_{\rm var}\gtrsim1$ & dynamic inversion \\
\bottomrule
\end{tabular*}
\setlength{\tabcolsep}{6pt}
\renewcommand{\arraystretch}{1.0}
\normalsize
\end{table*}

For the four benchmarks, the calibration ratio $Q_b/Q_s$ illustrates the range of requirements. A 1\% false-signal tolerance permits very loose background calibration for the photon-rich solar analog, $\epsilon_{\rm cal}\lesssim1.4\times10^{-1}$, but requires $\epsilon_{\rm cal}\lesssim9.8\times10^{-7}$ for the bright protoplanetary subfield because $Q_b/Q_s\simeq1.0\times10^4$. The white-dwarf and M87*-like cases are intermediate, with 1\% limits of $\epsilon_{\rm cal}\lesssim5.2\times10^{-3}$ and $8.7\times10^{-3}$, respectively. These numbers explain why high SSIM in a controlled scalar test should not be overread: for faint or high-background targets, calibration and background covariance can dominate the final science error even when photon rates are adequate.

Table~\ref{tab:scales} collects the numerical SGL scale constants and image-plane conversions used in the radiometric and observability estimates below.

\begin{table*}[t]
\caption{SGL scale constants and fiducial observatory conversions at $z=650\,\AU$.}
\label{tab:scales}
\begin{ruledtabular}
\begin{tabular}{lll}
Quantity & Value & Role in observability \\
\colrule
Solar Schwarzschild radius $r_g$ & $2.953\,\km$ & sets lensing phase and focal distance \\
Impact parameter $b$ & $1.089R_\odot$ & sets formal angular response \\
Formal response at $1\,\um$ & $0.103\,\nas$ & ideal diffraction scale, not delivered map scale \\
Ideal monopole gain at $1\,\um$ & $1.17\times10^{11}$ & upper amplification scale \\
$1\,\nas$ in image plane & $0.471\,\m$ & compact-source sampling scale \\
$1\,\uas$ in image plane & $0.471\,\km$ & white dwarfs, black-hole rings, AGN BLR \\
$1\,\mas$ in image plane & $471\,\km$ & nearby stellar disks, disk subfields \\
$1''$ in image plane & $4.71\times10^5\,\km$ & full disks, lens systems, galaxy fields \\
Retarget by $1^\circ$ & $11.3\,\AU$ transverse & target-access constraint \\
Solar irradiance & $3.22\times10^{-3}\,\mathrm{W\,m^{-2}}$ & power and thermal-background scale \\
30\% photovoltaic output & $9.66\times10^{-4}\,\mathrm{W\,m^{-2}}$ & solar power is negligible \\
\end{tabular}
\end{ruledtabular}
\end{table*}

\begin{table*}[t]
\caption{Representative scalar simulations.  $Q_s$ and $Q_b$ are annular source and residual-background rates under the finite-source-limited effective-gain model, generalizing the $Q_{\rm exo}$ and $Q_{\rm cor}$ notation of the reflected-light exoplanet benchmark \cite{Turyshev2026DirectHighResExo}.  The adopted $\SNRc^{\rm eff}$ is the effective convolved-raster information floor injected before deconvolution after imposing the photon, calibration, systematic, and PSF-mismatch terms of Eq.~(\ref{eq:snrc_general}); it is \emph{not} the photon-only SNR $\SNRc^{\rm phot}$ of Eq.~(\ref{eq:snrc_phot}), which for the bright cases is orders of magnitude larger (cf.\ Table~\ref{tab:self_luminous_rates}). The column $n$ is the linear raster dimension, so the total number of image-plane samples is $N=n^2$.  Wall dwell assumes the listed number of spacecraft and excludes slews, calibration overhead, repeated phase coverage, and downlink. \ $^{\dagger}$The M87*-like row uses demonstration millimetre gain (${\cal G}_{\rm eff}=10^4$) and background assumptions, not a validated mm-SGL receiver/plasma/occultation model; its metrics should be cited as assumed-gain information-recovery results only.}
\label{tab:sim_assumptions}
\begin{ruledtabular}
\begin{tabular}{lccccccccc}
Case & $n$ & $D_{\rm img}$ & $\Delta_{\rm img}$ & $t_{\rm samp}$ & $N_{\rm sc}$ & ${\cal G}_{\rm eff}$ & $Q_s$ & $Q_b$ & $\SNRc$ \\
 & & [km] & [m] & [s] & & & [$\s^{-1}$] & [$\s^{-1}$] & \\
\colrule
Solar $10\,\pc$ & 160 & 438.5 & 2740 & 2 & 8 & $6.9\times10^3$ & $8.7\times10^{10}$ & $6.2\times10^9$ & 120 \\
Magnetic white dwarf, $10\,\pc$ & 128 & 4.02 & 31.4 & 20 & 4 & $7.5\times10^5$ & $3.2\times10^9$ & $6.2\times10^9$ & 95 \\
M87* ring/jet base$^{\dagger}$ & 128 & 19.8 & 155 & 30 & 12 & $1.0\times10^4$ & $2.6\times10^{10}$ & $3.0\times10^{10}$ & 38 \\
Bright $0.1\,\AU$ protoplanet subfield & 128 & 337 & 2631 & 300 & 12 & $9.0\times10^3$ & $5.5\times10^6$ & $5.6\times10^{10}$ & 24 \\
\end{tabular}
\end{ruledtabular}
\end{table*}

\subsection{Photon-rate comparison: self-luminous targets versus reflected-light exoplanets}
\label{sec:self_luminous_rates}

A concise photon-rate comparison belongs in this paper because it tests the central claim that the SGL has broad value beyond exoplanet imaging.  A full yield calculation with target catalogs, solar latitude, wavelength-dependent coronagraph/occulter throughput, and mission trajectory should be a follow-up study.  Here we use a deliberately transparent estimate to compare the reflected-light exo-Earth benchmark with the self-luminous and high-surface-brightness targets discussed below.

The reference reflected-light case is an Earth analog at $30\,\pc$ in the optical/NIR SGL regime.  The companion direct-imaging assessment gives $m_V\simeq27.77$ and $f_p\simeq7.84\times10^{-4}\,{\rm phot\,m^{-2}\,s^{-1}\,nm^{-1}}$ for a full-phase Earth twin at $10\,\pc$; scaling to $30\,\pc$ gives $m_V\simeq30.2$ \cite{Turyshev2025DirectHighRes,Turyshev2026DirectHighResExo}.  In the resolved exoplanet-imaging notation the reflected planet and solar-corona rates are $Q_{\rm exo}$ and $Q_{\rm cor}$.  Here $Q_s$ and $Q_b$ denote the generalized SGL-coupled source and residual-background rates for arbitrary target classes; for the exo-Earth reference we set $Q_s\equiv Q_{\rm exo}=8.01\times10^4\,\s^{-1}$ and $Q_b\equiv Q_{\rm cor}=6.20\times10^9\,\s^{-1}$ for a $1\,\m$ aperture at $650\,\AU$. The value $Q_{\rm exo}$ is adopted directly from the companion reflected-light assessment \cite{Turyshev2025DirectHighRes,Turyshev2026DirectHighResExo} rather than recomputed from Eqs.~(\ref{eq:q_ab})--(\ref{eq:qs_eff}); it corresponds to a full reflected-light spectral integration with effective throughput--bandwidth product of order unity, not to the simplified continuum prescription ($\eta=0.25$, $\beta=0.1$) used for the self-luminous rows.  Applying the continuum prescription to the same $m_V\simeq30.2$ and $\mathcal{G}_{\rm eff}=2.3\times10^6$ inputs would instead yield $Q_s\simeq2.1\times10^3\,\s^{-1}$, smaller by the reflected-light integration factor; the two normalizations should not be compared directly.  The self-luminous rows of Table~\ref{tab:self_luminous_rates} are internally controlled from Eqs.~(\ref{eq:q_ab})--(\ref{eq:qs_eff}), whereas the exo-Earth reference row is a cross-normalization to the companion benchmark.

The photon-limited convolved-raster SNR and the corresponding dwell $t_{30}$ to reach $\SNRc^{\rm phot}=30$ are given by Eq.~(\ref{eq:snrc_phot}); for the reflected exo-Earth benchmark $\SNRc^{\rm phot}(1\,\s)\simeq1.0$ and $t_{30}\simeq8.7\times10^2\,\s$.  The $t_{30}$ column of Table~\ref{tab:self_luminous_rates} is computed from that relation. For the other optical/NIR entries in Table~\ref{tab:self_luminous_rates}, we use Eq.~(\ref{eq:qs_eff}) with $\eta=0.25$, $\Delta\nu/\nu=0.1$ unless otherwise stated, and the fiducial residual solar foreground
\begin{equation}
  Q_b\simeq6.2\times10^9
  \Big(\frac{d}{1\,\m}\Big)^2
  \Big(\frac{\Delta\nu/\nu}{0.1}\Big)\s^{-1}.
  \label{eq:fiducial_background}
\end{equation}
The long-wavelength entries are retained as concept-level estimates because their backgrounds, receiver temperatures, solar plasma, and ring-extraction physics are different from the optical/NIR case. 

\begin{figure*}[t]
  \centering
  \includegraphics[width=0.86\textwidth]{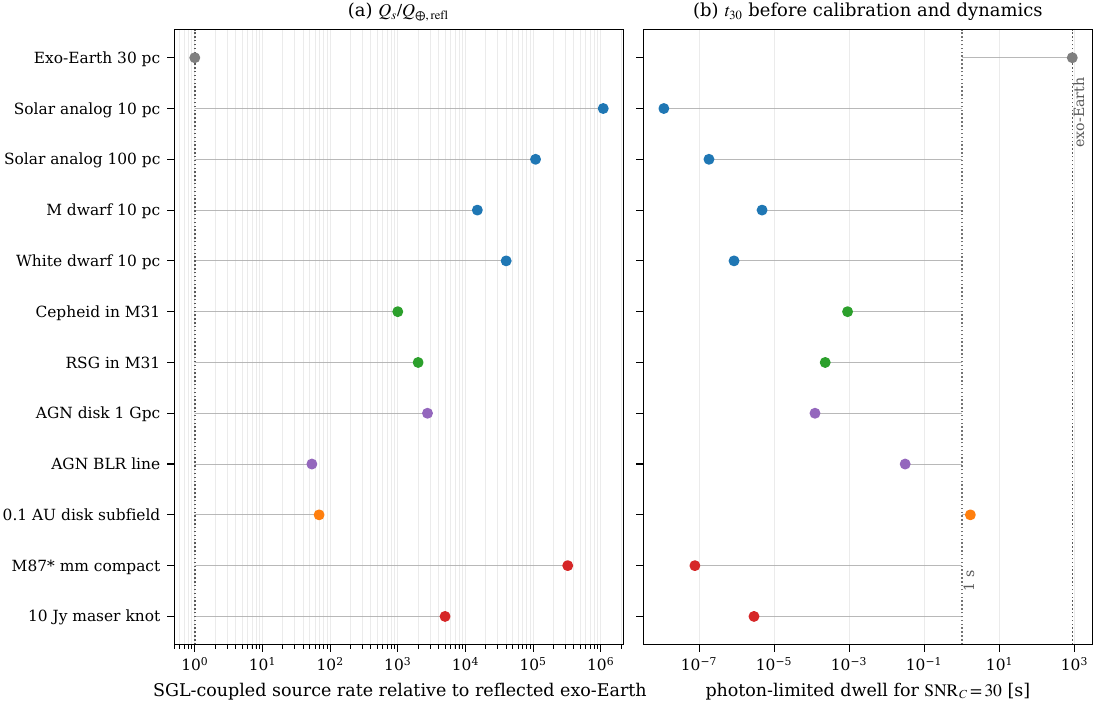}
  \caption{Photon-rate comparison for the target classes in Table~\ref{tab:self_luminous_rates}.  Panel (a) shows the SGL-coupled source rate relative to the reflected-light exo-Earth reference, $Q_{\rm exo}$.  Panel (b) shows the photon-limited dwell time required for $\SNRc=30$ before calibration covariance, detector dynamic range, slew, cadence, annular extraction, and reconstruction penalties.  Color encodes target class only; bar size has no additional physical meaning.  The main conclusion is limited but important: many self-luminous compact sources are not photon-starved in the SGL, whereas delivered performance is controlled by background subtraction, PSF knowledge, metrology, dynamic range, temporal variability, scan overhead, and access geometry.}
  \label{fig:self_luminous_rates}
\end{figure*}

Figure~\ref{fig:self_luminous_rates} visualizes the photon-rate comparison in Table~\ref{tab:self_luminous_rates} by plotting both the SGL-coupled source rate relative to the reflected exo-Earth reference and the corresponding photon-limited dwell for $\SNRc=30$.

\begin{table*}[t]
\caption{Order-of-magnitude photon-rate comparison for representative SGL target classes at $z=650\,\AU$.  The brightness column is an input flux descriptor: AB/V magnitude for optical/NIR continuum cases and flux density or line flux for long-wavelength examples.  $Q_s$ is the generalized SGL-coupled source rate after finite-source gain and the stated aperture/ring-throughput model; $Q_b$ is the residual background rate used in the planning estimate, reducing to $Q_{\rm cor}$ for the reflected exo-Earth reference.  $t_{30}$ is the photon-limited dwell to reach $\SNRc=30$ before calibration covariance, detector dynamic range, slew, reconstruction, cadence, and PSF-model penalties.  The M87* mm row is a long-wavelength demonstration assumption, not a validated receiver/background/plasma/occultation prediction.  The AGN BLR line-channel row uses a narrowband fractional bandwidth $\beta=\Delta\nu/\nu=0.01$ appropriate to a single $R_\lambda\sim100$ velocity channel, a factor of ten below the $\beta=0.1$ continuum default used for the other optical/NIR rows; applying the continuum $\beta$ to this row would overstate $Q_s$ by an order of magnitude.}
\label{tab:self_luminous_rates}
\setlength{\tabcolsep}{3.5pt}
\renewcommand{\arraystretch}{1.10}
\begin{ruledtabular}
\begin{tabular}{lcccccccc}
Target & Flux input & $d$ & $D_{\rm img}$ & ${\cal G}_{\rm eff}$ & $Q_s$ & $Q_b$ & $\SNRc^{\rm phot}(1\,\s)$ & $t_{30}$ \\
 & & [m] & [km] & & [$\s^{-1}$] & [$\s^{-1}$] & & [s] \\
\colrule
Exo-Earth $30\,\pc$ & $m_V\simeq30.2$ & 1 & 1.34 & $2.3\times10^6$ & $8.0\times10^4$ & $6.2\times10^9$ & $1.0$ & $8.7\times10^2$ \\
Solar $10\,\pc$ & $m\simeq4.83$ & 1 & 438 & $6.9\times10^3$ & $8.7\times10^{10}$ & $6.2\times10^9$ & $2.8\times10^5$ & $1.1\times10^{-8}$ \\
Solar $100\,\pc$ & $m\simeq9.83$ & 1 & 43.8 & $6.9\times10^4$ & $8.7\times10^9$ & $6.2\times10^9$ & $7.1\times10^4$ & $1.8\times10^{-7}$ \\
M dwarf $10\,\pc$ & $m\simeq11$ & 1 & 110 & $2.8\times10^4$ & $1.2\times10^9$ & $6.2\times10^9$ & $1.4\times10^4$ & $4.7\times10^{-6}$ \\
WD $10\,\pc$ & $m\simeq13.5$ & 1 & 4.02 & $7.5\times10^5$ & $3.2\times10^9$ & $6.2\times10^9$ & $3.3\times10^4$ & $8.1\times10^{-7}$ \\
Cepheid in M31 & $m\simeq20.4$ & 1 & 0.28 & $1.1\times10^7$ & $8.0\times10^7$ & $6.2\times10^9$ & $1.0\times10^3$ & $8.8\times10^{-4}$ \\
RSG M31 & $m\simeq16.4$ & 1 & 5.62 & $5.4\times10^5$ & $1.6\times10^8$ & $6.2\times10^9$ & $2.0\times10^3$ & $2.2\times10^{-4}$ \\
AGN disk $1\,{\rm Gpc}$ & $m\simeq18$ & 1 & 0.94 & $3.2\times10^6$ & $2.2\times10^8$ & $6.2\times10^9$ & $2.7\times10^3$ & $1.2\times10^{-4}$ \\
AGN BLR line channel & $m\simeq19$, $R_\lambda\sim100$ & 1 & 1.90 & $1.6\times10^6$ & $4.3\times10^6$ & $6.2\times10^8$ & $1.7\times10^2$ & $3.0\times10^{-2}$ \\
$0.1\,\AU$ disk field & $m\simeq18$ & 3 & 337 & $9.0\times10^3$ & $5.5\times10^6$ & $5.6\times10^{10}$ & $23$ & $1.7$ \\
M87* mm source & $0.7\,{\rm Jy}$ & 3 & 19.8 & $1.0\times10^4$ & $2.6\times10^{10}$ & $3.0\times10^{10}$ & $1.1\times10^5$ & $7.5\times10^{-8}$ \\
Compact maser knot & $10\,{\rm Jy}$ line & 10 & 10 & $1.0\times10^3$ & $4.0\times10^8$ & $1.0\times10^8$ & $1.8\times10^4$ & $2.9\times10^{-6}$ \\
\end{tabular}
\end{ruledtabular}
\renewcommand{\arraystretch}{1.0}
\end{table*}

The values in Table~\ref{tab:self_luminous_rates} should be interpreted as common-scale photon-rate benchmarks, not as final exposure-time predictions.  They intentionally use a simplified foreground prescription to compare target classes on the same footing.  A flight exposure-time calculator must replace $Q_b$ and $\eta_{\rm ann}$ with wavelength-, solar-latitude-, aperture-, occultation-, detector-, ring-extraction-, and target-environment-specific models.  The very small values of $t_{30}$ for bright self-luminous targets indicate only that photon statistics are unlikely to be the limiting term once the SGL-coupled source rate is high; they do not imply that microsecond spacecraft dwells are operationally useful.

The estimates show why a general SGL observatory is not merely an exoplanet instrument.  Stellar surfaces, white dwarfs, Local Group supergiants and Cepheids, compact AGN disks, and selected long-wavelength compact sources deliver source rates tens to millions of times larger than the reflected-light exo-Earth reference under the planning assumptions.  The scientific implication is not that the observations require ultra-short dwells, but that high source rates can support repeated sampling, phase registration, spectroscopy, polarimetry, and dynamic tomography once calibration, PSF knowledge, ring extraction, metrology, and focal-line access are solved. The Einstein-ring signal may also support target acquisition and closed-loop ring centroiding, but it does not relax PSF, metrology, background, or focal-line constraints.

The photon advantage does not remove the focal-line access constraint.  Observing many unrelated targets still requires large transverse motion or a distributed focal-shell architecture.  It does, however, make multiple-target campaigns more plausible once a spacecraft is near a chosen line or cluster of lines: dwell time is no longer the dominant cost for many self-luminous targets, and the campaign is controlled instead by pointing, scanning, calibration, target variability, downlink, and retargeting geometry.

\begin{figure*}[t]
  \centering
  \includegraphics[width=0.86\textwidth]{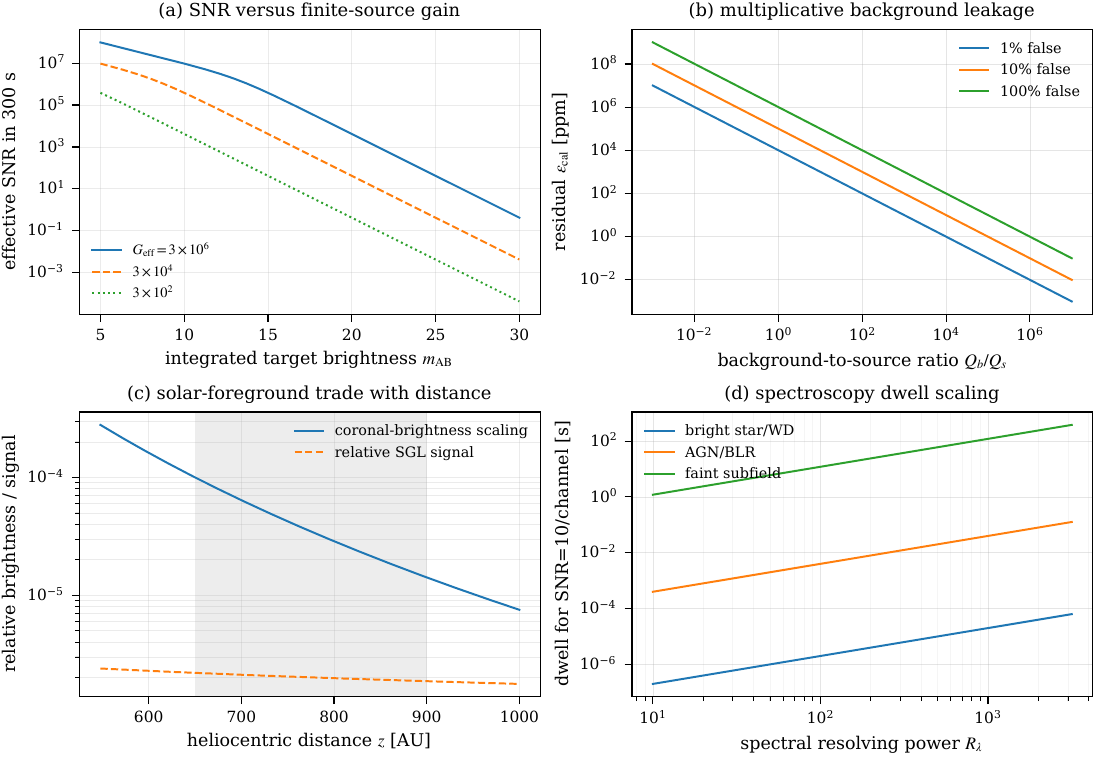}
  \caption{Radiometric, foreground, and spectral scaling diagnostics.  Panel (a) shows effective SNR in a 300 s dwell as a function of integrated AB magnitude for representative finite-source-limited gains.  Panel (b) shows how multiplicative residuals on a bright background become false source signal, $\epsilon_{\rm cal}Q_b/Q_s$.  Panel (c) illustrates the distance trade between a steep coronal-brightness scaling and an extended-source SGL signal scaling; detailed values require a wavelength-, solar-latitude-, and instrument-specific foreground model.  Panel (d) gives representative spectral-channel dwell-time scalings.  These are planning curves, not an end-to-end instrument budget.}
  \label{fig:radiometry}
\end{figure*}

Figure~\ref{fig:radiometry} gives the companion planning scalings for magnitude, calibration leakage, heliocentric-distance foreground tradeoffs, and spectral-channel dwell time.

The operational consequences are target-class dependent.  For bright stellar, white-dwarf, Cepheid, and AGN targets, the image-plane sampling rate can be limited by spacecraft motion, metrology update rate, detector full well, and ring-extraction stability rather than by photon statistics.  Pixel dwell can be shortened to avoid saturation and then repeated many times to build dynamic maps.  Spectroscopy can be traded against cadence: $R_\lambda\sim10^2$--$10^3$ channels become plausible for many bright targets, while very high-resolution spectropolarimetry should be limited to selected lines or selected image-plane samples.  Target acquisition is also easier than for reflected-light exoplanets, because the high-SNR annular signal can support ring-centroiding and closed-loop scans.  Navigation tolerance, however, is not relaxed by photon rate; it remains set by the desired delivered spatial resolution, $\sigma_\rho\lesssim\epsilon\Delta_{\rm img}=\epsilon D_{\rm img}/n$, and by the accuracy with which the measured coordinates are included in the forward operator.  Finally, high photon rates enable cadence: repeated scans can follow stellar rotation, white-dwarf spin, pulsation phase, AGN reverberation, or jet variability, provided the temporal inverse model is part of the observing plan.

\section{Representative simulation results}
\label{sec:sim_results}

\begin{figure*}[h!]
  \centering
  \includegraphics[width=0.86\textwidth]{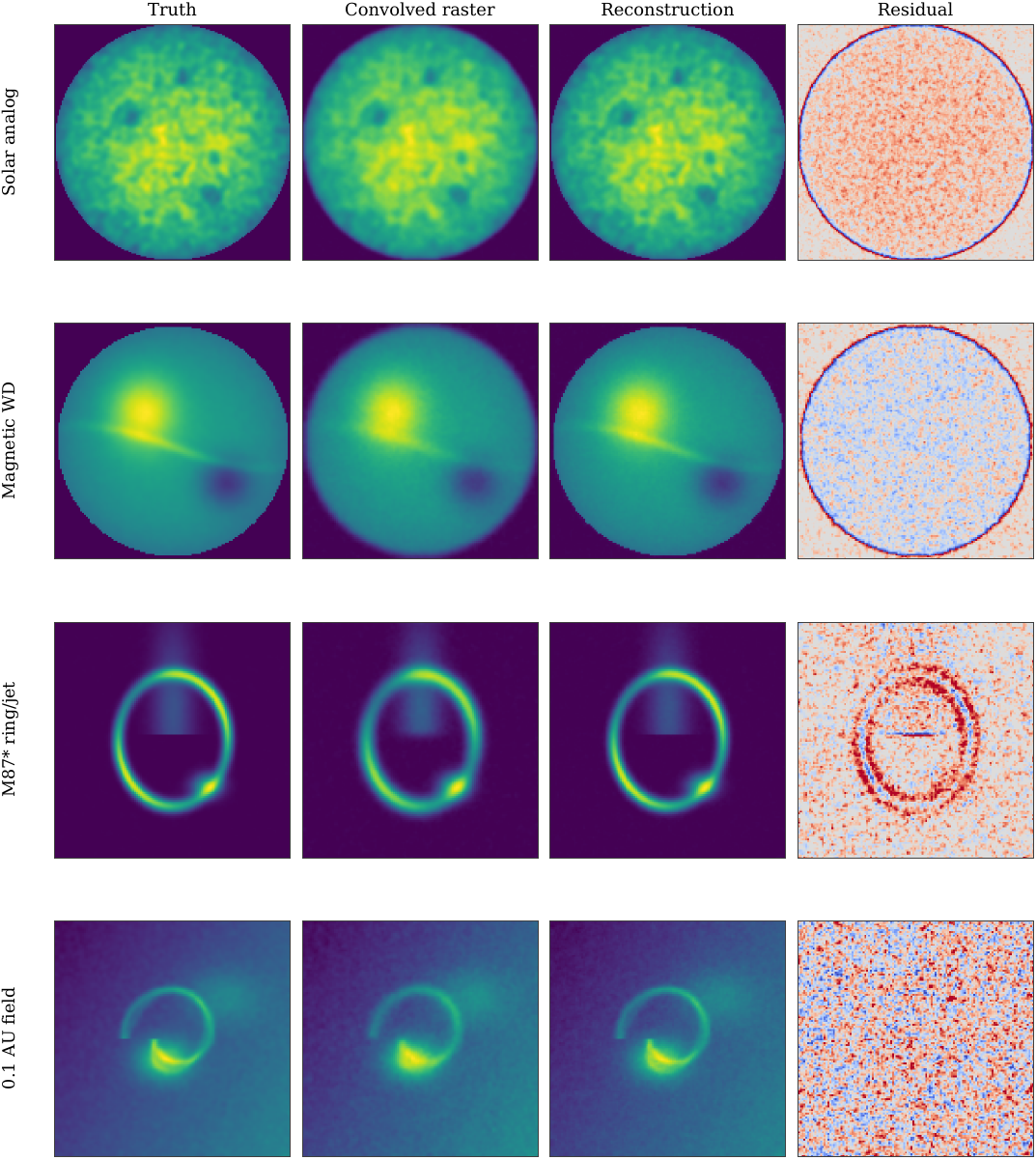}
  \caption{Representative scalar SGL benchmark simulations. Each row shows the input truth, SGL-convolved raster with noise and structured residuals, Wiener/Tikhonov reconstruction using the nominal monopole kernel, and residual map. Display stretches are chosen for morphology; all quantitative values are reported separately in Table~\ref{tab:metrics}. The simulations use the parameters in Table~\ref{tab:sim_assumptions} and the metric definitions in Sec.~\ref{subsec:metrics}. The figure demonstrates scalar information recovery under the stated benchmark protocol: prescribed support, fixed regularization, imposed effective $\SNRc$ floor, and smooth low-order PSF/background residuals.  Coherent ring-correlated structure visible in the M87*-like residual panel is the expected signature of the imposed low-order kernel mismatch rather than photon noise,  and is the morphological counterpart of the $f_J$ sensitivity quantified in Fig.~\ref{fig:wd_j_multipole}.}
  \label{fig:simulation_gallery}
\end{figure*}

Figure~\ref{fig:simulation_gallery} gives the truth images, SGL-convolved rasters, reconstructed images, and residuals.  Table~\ref{tab:metrics} gives the corresponding fixed-protocol scalar recovery metrics.  The protocol is common across the four rows: analytic scene, prescribed science support, stationary scalar aperture-averaged kernel, imposed effective $\SNRc$ floor, specified structured residuals, non-negativity/support constraints, and one regularization convention.  Under that protocol, the reported $\SSIM$, NRMSE, contrast, $\FRC_{50}$, and $f_{\rm out}$ values show that the selected morphologies survive SGL convolution and deconvolution with scientifically useful spatial information.  Figure~\ref{fig:sensitivity_boundaries}(b) gives the associated information-floor sensitivity by varying $\SNRc$ while holding the scene, support, kernel family, and regularization fixed.

\paragraph*{Static and dynamic regimes.}
The reconstructions in Fig.~\ref{fig:simulation_gallery} are frozen-scene benchmarks.  This is the correct entry test for scalar recoverability; time dependence becomes part of the inverse problem when $\chi_t=T_{\rm wall}/t_{\rm var}\gtrsim0.1$--1.  The white-dwarf case is closest to the static limit because $D_{\rm img}=4.02\,\km$ and many targets can be phase-folded on spin or pulsation.  Stellar surfaces, active atmospheres, accretion structures, black-hole hot spots, AGN reverberation, and protoplanetary subfields require $O(\boldsymbol{\xi},t)$ rather than a single map.  For these classes the static metrics establish spatial recoverability; the delivered science product is a dynamic posterior using Eq.~(\ref{eq:joint_inverse}).

The stellar-surface simulation is photon-rich and calibration-limited. A $160\times160$ map of a solar analog at $10\,\pc$ has image-plane pitch $\Delta_{\rm img}=2.74\,\km$ in the image plane and a source-plane pixel scale of $8700\,\km$.  With an adopted $\SNRc=120$ and a small unmodeled PSF perturbation, the reconstruction recovers starspots, plage structure, large-scale granulation, and limb darkening with $\SSIM=0.993$, $\mathrm{NRMSE}=0.044$, and FRC$_{50}$ $=1.7\times10^4\,\km$.  The result is not limited by formal SGL resolution.  It is limited by scan time, stellar evolution, detector dynamic range, and model calibration.

The white-dwarf simulation is the most favorable compact-imaging case.  A $10\,\pc$ Earth-size white dwarf projects to only $4.02\,\km$, so a $128\times128$ raster has $31\,\m$ spacing and $100\,\km$ source pixels.  With $\SNRc=95$, mild PSF mismatch, and structured residuals, the recovered surface has $\SSIM=0.918$, $\mathrm{NRMSE}=0.091$, and FRC$_{50}$ $=315\,\km$.  The recovered hot and cool regions are scientifically meaningful at scales far below any conventional optical facility.  Spectropolarimetric channels would turn such maps into constraints on magnetic-field topology and accretion belts.

Because the white-dwarf benchmark has the smallest raster pitch, $\Delta_{\rm img}=31.4\,\m$, it is the sharpest transfer-function stress test in the present set.  We therefore added the $J_2/J_4$-anchored closure calculation defined by Eqs.~(\ref{eq:j_proxy_kernel})--(\ref{eq:j_proxy_scale}).  On the same $128\times128$ grid and with the same effective $\SNRc=95$, the auxiliary white-dwarf scene gives $\SSIM=0.933$ and $\mathrm{NRMSE}=0.074$ when truth and inverse both use the nominal monopole scalar kernel.  With a residual low-order transfer-function error $f_J=0.05$, corresponding to $14\,\m=0.45\Delta_{\rm img}$, the reconstruction remains morphologically stable with $\SSIM=0.852$ and $\mathrm{NRMSE}=0.144$.  With the full quadrupole-scale residual left unmodeled, $f_J=1$ and $a_2=281\,\m\simeq9\Delta_{\rm img}$, the same inverse degrades to $\SSIM=0.628$ and $\mathrm{NRMSE}=0.294$.

\begin{figure*}[t]
  \centering
  \includegraphics[width=0.90\textwidth]{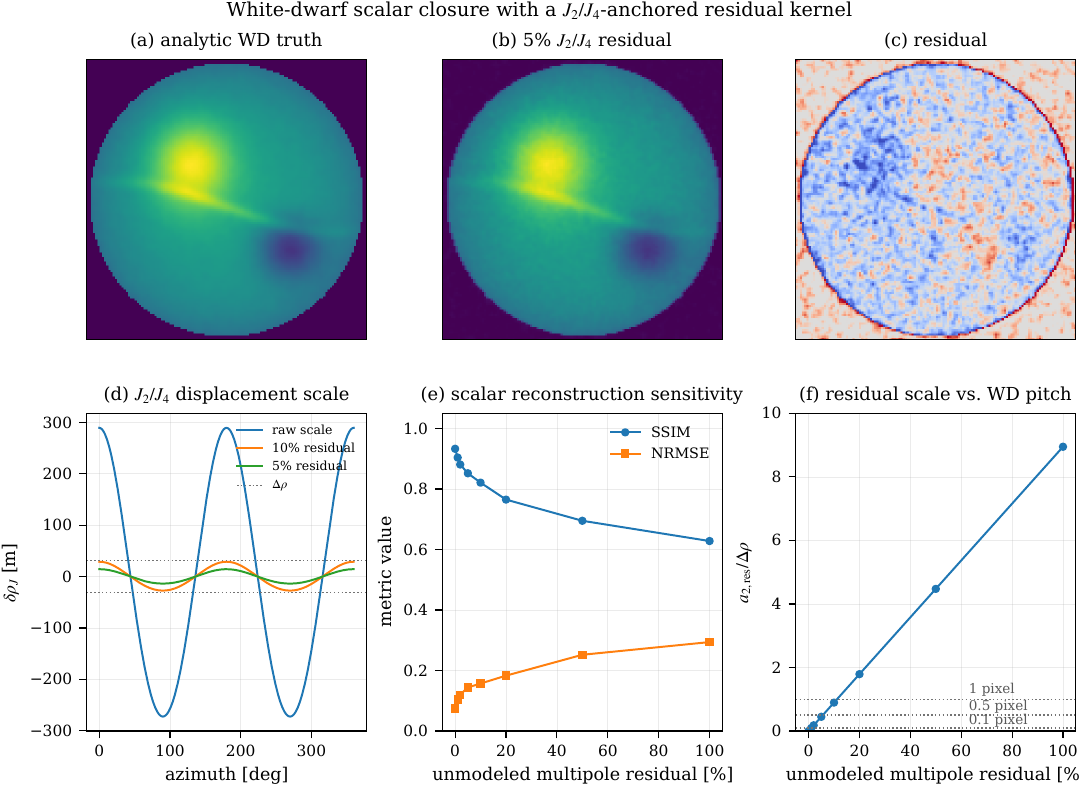}
  \caption{White-dwarf scalar closure test with a low-order solar-multipole residual surrogate. Panels (a)--(c) show an analytic white-dwarf truth image, the reconstruction when the truth kernel contains a 5\% residual of the $J_2/J_4$ displacement scale, and the corresponding residual map. Panel (d) shows the raw and residual image-plane displacement scale from Eqs.~(\ref{eq:j_proxy_kernel})--(\ref{eq:j_proxy_scale}); the raw quadrupole-scale term is much larger than the $31.4\,\m$ white-dwarf raster pitch. Panel (e) shows scalar reconstruction metrics versus the unmodeled residual fraction. Panel (f) expresses the residual quadrupole scale in units of the white-dwarf image-plane pitch. The surrogate is anchored to plausible zonal-harmonic image-plane scales, but it is not a wave-optical solar-multipole, plasma, or extended-Sun PSF and should not be interpreted as a literal radial displacement of the true SGL response. The result is a controlled closure test: calibrated low-order residuals are tolerable in the scalar benchmark, while an unmodeled solar-quadrupole-scale distortion is not.}
  \label{fig:wd_j_multipole}
\end{figure*}

The important point is operational.  The $f_J\rightarrow1$ regime is a failure bound, not an observing mode.  A quantitative white-dwarf SGL observation requires the deterministic low-order solar response to be calibrated down to the science-contrast scale, which in this benchmark corresponds approximately to a residual displacement $\lesssim0.5\Delta_{\rm img}$.  Figure~\ref{fig:wd_j_multipole} shows that the scalar inverse is stable at that residual scale and unstable when a raw quadrupole-scale distortion remains in the data.  The scalar displacement surrogate is not a wave-optical solar-multipole/plasma PSF; its role is to translate the physical transfer-function problem into a measurable image-plane residual requirement for the PSF library represented by Eq.~(\ref{eq:psf_library}). 

The M87*-like row is an assumed-gain millimetre information-recovery test.  It establishes the conditional statement that, for a compact $42\,\mu$as ring/jet source projecting to $D_{\rm img}=19.8\,\km$, a scalar SGL measurement with ${\cal G}_{\rm eff}=10^4$, residual background $Q_b=3.0\times10^{10}\,\mathrm{s^{-1}}$, and effective $\SNRc=38$ retains sub-$\mu$as structure after deconvolution.  Under those inputs the benchmark recovers the ring, asymmetry, hot spot, and jet-base extension with $\SSIM=0.973$ and $\FRC_{50}=0.66\,\mu$as.  The confidence-building result is therefore geometric and information-theoretic: M87*-scale structure is compact in the SGL image plane and recoverable at the specified scalar information quality.  The separate technology closure is mm-specific receiver gain, solar thermal background, plasma phase, scattering, occultation, annular coupling, polarization leakage, dynamic imaging.  M87* is the natural first compact-object target because its variability is slower than Sgr~A*; it remains a long-wavelength SGL observatory mode, not an optical/NIR extrapolation.

The protoplanetary subfield simulation demonstrates the correct target selection logic.  Full disks are too large in the image plane, but a $0.1\,\AU$ subfield at $140\,\pc$ projects to $337\,\km$.  The scalar benchmark recovers a circumplanetary disk, spiral wake, and gap-edge structure at $\FRC_{50}=1.6\times10^{-3}\,\AU$ with $\SSIM=0.923$.  The source is background-limited in the optical/NIR planning model, and the required wall dwell is several days even with 12 spacecraft.  The scientific product would be a targeted image of a known protoplanetary environment, not a survey image of an entire disk.

\begin{table*}[t]
\caption{Scalar reconstruction metrics for the four benchmark scenes. Metrics are computed on the support masks defined in Sec.~\ref{sec:simulation_method}. $\FRC_{50}$ is reported in the natural source unit: kilometres on stellar and white-dwarf surfaces, microarcseconds for the M87*-like source, and AU for the protoplanetary subfield. It is a truth-referenced benchmark resolution diagnostic. The reported $\SNRr$ is a masked post-reconstruction residual statistic and is not expected to equal the unconstrained conditioning estimate in Eq.~(\ref{eq:snr_penalty}). The $\SSIM$ values are fixed-protocol recovery metrics conditioned on the adopted $\SNRc$ floors, kernels, masks, and regularization; Fig.~\ref{fig:sensitivity_boundaries}(b) gives the corresponding information-floor sensitivity. Values are reported for one fixed-seed realization per case, with deterministic uncertainty dominated by kernel mismatch, support choice, effective floor, and regularization. A target-specific performance study should promote this table to ensemble means and standard deviations over those variables.}
\label{tab:metrics}
\begin{ruledtabular}
\begin{tabular}{lccccccc}
Case & $\SNRc$ & $\SNRr$ & SSIM & NRMSE & contrast & $\FRC_{50}$ & $f_{\rm out}$ \\
\colrule
Solar analog surface & 120 & 120 & 0.993 & 0.044 & 0.999 & $1.7\times10^4\,\km$ & 0.006 \\
Magnetic white dwarf & 95 & 26.6 & 0.918 & 0.091 & 0.996 & $315\,\km$ & 0.018 \\
M87* ring/jet base & 38 & 72.4 & 0.973 & 0.0148 & 1.000 & $0.66\,\uas$ & 0.021 \\
$0.1\,\AU$ protoplanet subfield & 24 & 18.1 & 0.923 & 0.087 & 0.996 & $1.6\times10^{-3}\,\AU$ & field \\
\end{tabular}
\end{ruledtabular}
\end{table*}

\section{SGL spectroscopy and spectropolarimetry}
\label{sec:spectroscopy}

The SGL science case is stronger when imaging is treated as a spatially resolved spectroscopy problem rather than as broadband map recovery alone.  A general data product is a cube, or time-dependent cube,
\begin{equation}
  I(\boldsymbol{\xi},\lambda,p,t),
\end{equation}
where $\boldsymbol{\xi}$ is source-plane position, $\lambda$ is wavelength, $p$ denotes polarization state, and $t$ is time.  The SGL does not measure this cube directly.  It measures annular or sector-integrated Einstein-ring fluxes while the spacecraft samples the parity-reversed image plane.  A spectral channel therefore has the same mapping and inverse problem as the broadband case,
\begin{equation}
 y_{i\ell ap}\sim {\rm Poisson}\!\Big[
 \int K_{i\ell ap}(\boldsymbol{\xi},t;\boldsymbol{\eta})
 I(\boldsymbol{\xi},\lambda_\ell,p,t)\,d^2\xi
 +b_{i\ell ap}\Big],
 \label{eq:spectral_measurement}
\end{equation}
but with a wavelength-dependent PSF, gain, corona, detector response, and astrophysical morphology.

\begin{figure*}[t]
  \centering
  \includegraphics[width=0.86\textwidth]{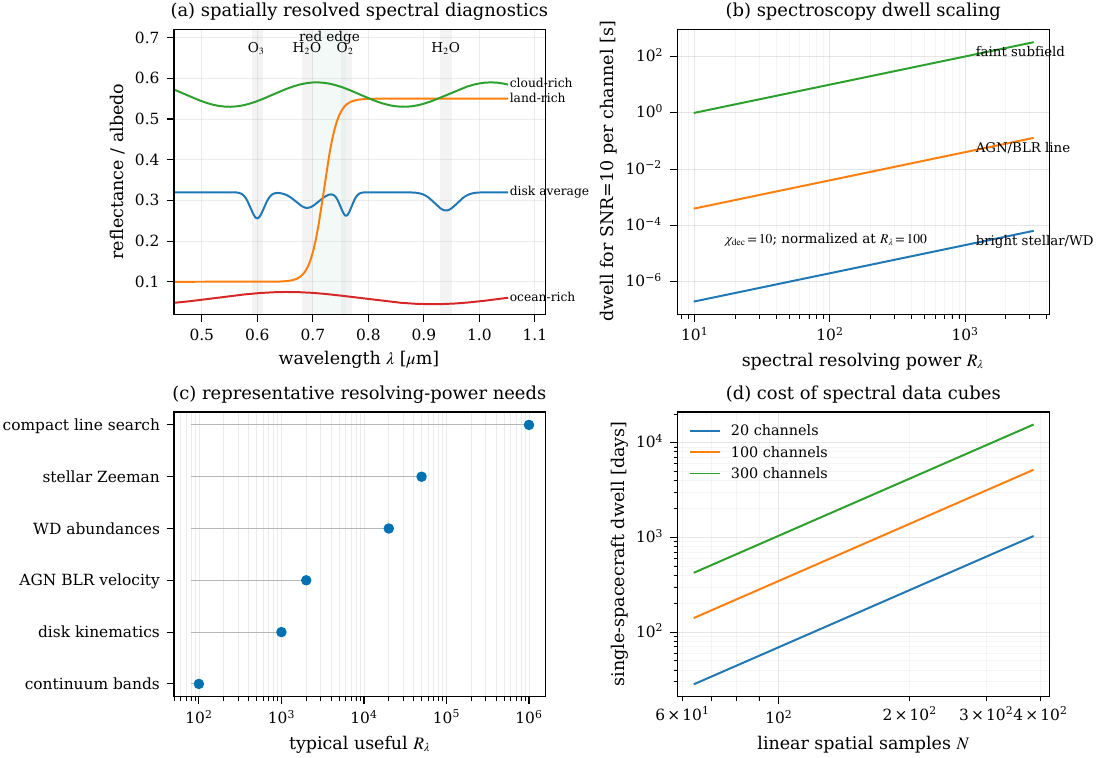}
  \caption{SGL spectroscopy diagnostics.  Panel (a) shows representative spatially resolved reflected-light spectra and molecular/surface intervals to illustrate why spatial context matters.  Panel (b) gives a photon-limited dwell-time scaling for SNR=10 per spectral channel for three representative brightness classes under a finite-source-limited gain model.  Panel (c) summarizes typical resolving-power regimes for several SGL science products.  Panel (d) shows the dwell-time cost of building spectral data cubes.  The curves are planning scalings, not end-to-end instrument simulations.}
  \label{fig:spectroscopy}
\end{figure*}

Figure~\ref{fig:spectroscopy} summarizes the corresponding spectroscopy diagnostics: representative spatially resolved spectra, photon-limited channel dwell, resolving-power regimes, and the dwell-time cost of spectral cubes.

For continuum spectroscopy, the per-channel source rate scales approximately as
\begin{equation}
  Q_{s,\ell}\simeq
  Q_{\rm tel}(m_{\rm AB})\,
  \eta_{{\rm ann},\ell}
  \min\Big[\bar\mu_{d,\ell},\frac{4b}{D_{{\rm img},\ell}},\mu_{0,\ell}\Big]
  \frac{\Delta\nu_\ell/\nu}{0.1},
  \label{eq:spectral_rate}
\end{equation}
so that increasing resolving power $R_\lambda=\lambda/\Delta\lambda$ reduces the photon rate per channel roughly as $R_\lambda^{-1}$ at fixed source brightness.  For a target product requiring SNR$_\ell$ in each reconstructed spatial/spectral element, a useful photon-limited dwell estimate is
\begin{equation}
  t_\ell \simeq \chi_{\rm dec}\,\SNR_\ell^2
  \frac{Q_{s,\ell}+Q_{b,\ell}}{Q_{s,\ell}^2},
  \label{eq:spectral_dwell}
\end{equation}
where $\chi_{\rm dec}\ge1$ is the reconstruction and spectral-extraction penalty.  Equation~(\ref{eq:spectral_dwell}) is only a planning relation; a flight design must propagate the full covariance of the coupled spectral inverse problem.

Spatially resolved SGL spectroscopy enables products that conventional facilities usually infer indirectly.  Stellar and white-dwarf applications include line-depth maps, abundance spots, Zeeman splitting, Stokes-parameter maps, Doppler velocity fields, and flare-localized spectra.  Planet-forming applications include line-channel maps of accretion shocks, molecular emission, scattered-light color, dust temperature, and kinematic perturbations around protoplanets.  Compact-object and AGN applications include velocity-resolved BLR maps, hot-spot spectra, jet-base spectral-index maps, polarization rotation, reverberation lags, and disk temperature profiles.  The common feature is that the SGL supplies source-plane sampling while spectroscopy supplies physical diagnostics; together they move the data product from ``a picture'' to a calibrated physical map.

The observing cost is substantial.  A cube with $N=n^2$ spatial samples and $N_\lambda$ spectral channels requires approximately
\begin{equation}
  T_{\rm dwell}\simeq \frac{N N_\lambda t_{\ell}}{N_{\rm sc}}
  =\frac{n^2 N_\lambda t_{\ell}}{N_{\rm sc}},
  \label{eq:cube_time}
\end{equation}
before slew, calibration, repeat visits, phase coverage, and downlink. This favors sparse spectral strategies: broadband maps for morphology, selected medium-resolution channels for molecular or kinematic diagnostics, and high-resolution spectroscopy only for bright compact targets or for a small number of spatial samples.  Multi-spacecraft sampling is especially valuable because it reduces temporal aliasing and permits repeated spectral measurements at the same image-plane locations.

\section{Quantitative science-case assessment}
\label{sec:science_cases}

\subsection{Resolved stellar surfaces, atmospheres, activity, and magnetism}

The unique SGL product is not merely a resolved stellar disk, but a time-dependent atlas of surface brightness, line depth, velocity, and polarization.  Such maps would localize active longitudes, measure spot and facula contrasts without light-curve degeneracies, separate photospheric contamination from exoplanet spectra, and provide spatial priors for stellar dynamo models \cite{Berdyugina2005,DonatiLandstreet2009}.  Conventional interferometers can image a few large nearby stars; the SGL would extend surface mapping to ordinary nearby dwarfs and subgiants if variability can be modeled.

Nearby stellar surfaces are a top-tier SGL science case.  A solar analog at $10\,\pc$ has angular diameter $0.93\,\mas$ and image-plane diameter $438\,\km$.  A $160\times160$ raster provides $8700\,\km$ source pixels; a $512\times512$ raster would provide $2700\,\km$ pixels but with a much larger scan burden.  The scientific return includes starspots, faculae, large-scale granulation, limb darkening, differential rotation, active-region emergence, flare sites, and magnetic topology from spectropolarimetry.  Optical interferometers such as CHARA have resolved the largest and nearest stellar disks at sub-milliarcsecond scales \cite{CHARA2005,Monnier2007,Roettenbacher2016}, but the SGL would move stellar imaging from low-order surface structure on special targets to high-resolution mapping of ordinary nearby stars.

The limiting factors are scan time and variability.  Solar-like granulation changes on minutes, active regions evolve on hours to days, and rotation changes the projected surface during a long raster.  A stellar SGL mission must therefore use rapid sampling, repeated phase coverage, and dynamic inversion.  The most useful observables are persistent maps and time-dependent parameter fields, not one static image.  Brightness is not the limitation; dynamic range and calibration are.

\subsection{White dwarfs and compact stellar remnants}

For white dwarfs, the SGL uniquely changes unresolved spectropolarimetric inference into spatially resolved surface physics \cite{WickramasingheFerrario2000,Ferrario2015}.  It could map temperature patches, Zeeman-split magnetic regions, accretion belts, metal-pollution spots, and rotation-dependent limb darkening on scales of hundreds of kilometres.  These measurements would constrain magnetic-field geometry, crystallization and cooling physics, accretion from debris disks, and atmospheric composition in a way not achievable with foreseeable optical interferometry.

White dwarfs are arguably the strongest non-exoplanet SGL targets.  An Earth-size white dwarf at $10\,\pc$ subtends $8.5\,\uas$ and projects to $4.0\,\km$.  A $128\times128$ raster samples $31\,\m$ in the image plane and about $100\,\km$ on the surface.  This is a near-ideal match: compact enough for efficient scanning, large enough to image rather than treat as a point source, self-luminous, and astrophysically rich.

The science return includes temperature maps, magnetic spots, Zeeman-resolved field topology, accretion belts in polluted or accreting white dwarfs, rotation, limb darkening, and surface abundance inhomogeneity.  These quantities are currently inferred from unresolved light curves, spectra, and spectropolarimetry.  The SGL would convert them into resolved surface maps.  The main limitations are PSF knowledge, sub-metre image-plane metrology for high spatial resolution, detector dynamic range, and spectropolarimetric calibration.  This case is near-term plausible for an optical/NIR SGL observatory and belongs in the highest-priority tier of the non-exoplanet portfolio.

Two phase-resolved extensions further strengthen this case.  Pulsating white dwarfs have coherent periods of order $10^2$--$10^3\,{\rm s}$, so repeated fast scans can in principle be phase-folded to recover surface eigenfunction patterns rather than only time-averaged temperature maps \cite{WingetKepler2008}.  Accreting magnetic white
dwarfs and polars provide a complementary periodic problem: hot spots, accretion curtains, and Zeeman-resolved magnetic regions could be mapped through the spin or orbital cycle, provided the dynamic inverse model is included in the observing plan \cite{Ferrario2015}.

\subsection{Protoplanetary disks, planet-forming regions, and circumstellar structure}

The SGL does not need to observe an entire disk to be scientifically decisive.  A conventional observatory can identify a protoplanet, gap edge, spiral arm, dust trap, or accretion shock; the SGL can then scan a selected image-plane subfield to recover morphology and spectra at sub-AU or circumplanetary scales.  A full disk can in principle be scanned, but the required image-plane area, dwell time, and dynamic modeling are much larger.  The strongest products are therefore targeted subfield images, line-channel maps, scattered-light colors, and thermal or molecular diagnostics tied to known planet-forming structures.

\begin{figure*}[t]
  \centering
  \includegraphics[width=0.86\textwidth]{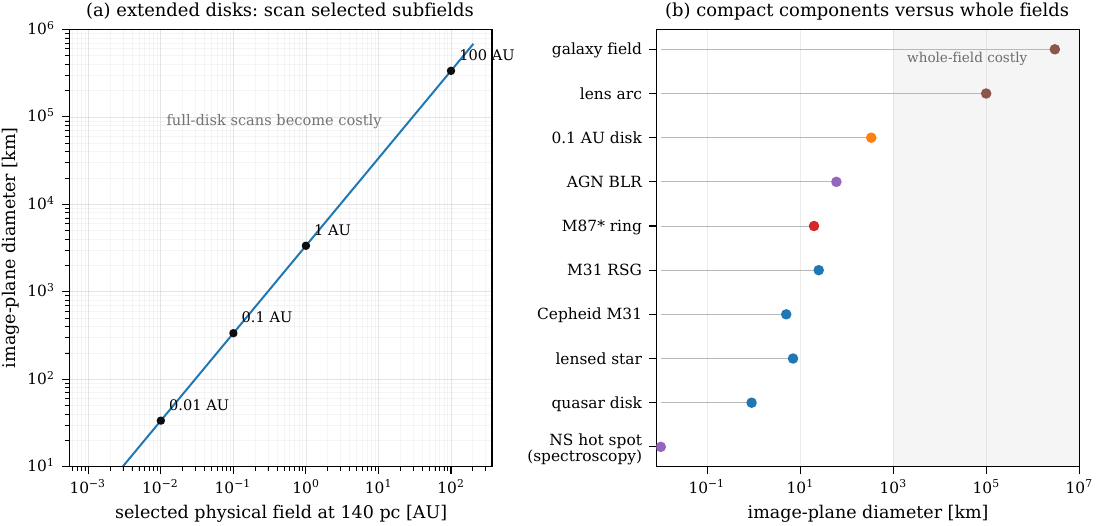}
  \caption{Subfield and compact-component constraints.  Panel (a) shows image-plane diameter for selected physical fields in a disk at $140\,\pc$; full disks are scan-prohibitive, while $0.01$--$0.1\,\AU$ subfields are plausible.  Panel (b) compares compact components and whole-field examples, adding a neutron-star hot-spot spectroscopy point, M87*-scale ring, AGN BLR, selected disk subfield, lens arc, and galaxy field. The SGL favors compact components and selected subfields, not arcsecond-scale systems.}
  \label{fig:subfield_compactness}
\end{figure*}

Figure~\ref{fig:subfield_compactness} shows why selected compact subfields are favored over full disks or arcsecond-scale systems: the SGL remains powerful for compact components, while whole-field image-plane diameters rapidly become scan-prohibitive.

The SGL is not a full protoplanetary-disk imager.  A $100\,\AU$ disk at $140\,\pc$ subtends $0.71''$ and projects to $3.4\times10^5\,\km$ in the image plane.  Full-disk SGL mapping at high resolution is operationally unattractive.  The strong science case is instead targeted subfield imaging around known structures: protoplanets, circumplanetary disks, Hill spheres, accretion shocks, gap edges, spiral wakes, dust traps, and inner-rim structures.  At $140\,\pc$, $0.1\,\AU$ projects to $337\,\km$ and $0.01\,\AU$ projects to $33.7\,\km$.

ALMA has revolutionized disk imaging at millimetre wavelengths, with surveys such as DSHARP resolving rings and gaps at a few-au scale \cite{ALMAPartnership2015,Andrews2018}.  The SGL is unique only if it addresses sub-AU or circumplanetary scales that ALMA, ELTs, JWST, or future interferometers cannot access.  This favors preselected subfields with known ephemerides.  The limitations are severe: host-star leakage, disk surface brightness, optical versus millimeter tracer differences, source variability from accretion, and the need for molecular-line or scattered-light spectroscopy.  This is long-term plausible for selected subfields, not a first-generation general disk survey.

Circumstellar structures around evolved stars occupy a related middle ground.  Stellar photospheres and inner molecular/dust-formation zones may be feasible, while full envelopes and bow shocks often become too large.  Good targets include AGB stars, red supergiants, Wolf-Rayet winds, dust-condensation regions, binary-induced spirals, and mass-transfer streams.  The SGL would be most valuable for the compact inner wind where convection, pulsation, dust formation, and outflow launching meet.

\subsection{Compact objects and their environments}

For compact objects, the SGL's value is the ability to sample structure below the angular scale of Earth-diameter VLBI, provided the relevant emission is observable in a practical wavelength band.  Potential measurements include photon-ring substructure, hot-spot trajectories, jet-base collimation, polarization morphology, and time-dependent accretion asymmetry.  These are parameter-estimation products as much as images: ring diameter, asymmetry, inclination, spin-sensitive morphology, and magnetic-field orientation can be inferred jointly from dynamic spatial and spectral measurements.

Black-hole and compact-object science is one of the highest-value long-term SGL applications.  M87* has an EHT ring diameter of order $42\,\uas$, and Sgr A* has a ring diameter of order $52\,\uas$ \cite{EHT2019IV,EHT2022SgrA}.  These project to $20$--$25\,\km$ at $650\,\AU$, a highly manageable image-plane scale.  The formal SGL response at $1.3\,\mathrm{mm}$ is $0.13\,\uas$, suggesting access to substructure within the photon ring, jet-launching region, and polarized emission morphology.

Consequently, the M87*-like benchmark should be read as a compact-source information-recovery test at millimetre angular scales, not as evidence that an optical/NIR SGL spacecraft can perform black-hole imaging without a separate long-wavelength receiver, occultation, plasma, and solar-background design.

The caveat is that black-hole shadow science is not naturally optical.  It requires mm/sub-mm instrumentation, occultation, solar thermal-background control, plasma modeling, polarization calibration, and dynamic imaging.  Sgr A* evolves on minute-to-hour timescales and likely requires simultaneous multi-spacecraft sampling.  M87* evolves more slowly and is the more plausible first compact-object SGL target.  The SGL could in principle move beyond horizon-scale detection toward precision tests of strong gravity, magnetic-field morphology, hot-spot dynamics, and jet launching, but this case is long-term plausible rather than near-term.

Neutron stars and stellar-mass black holes are more credible as amplified time/spectral targets than as resolved-imaging targets.  For example, a $25\,{\rm km}$ neutron star at $123\,{\rm pc}$ subtends only $\Theta\simeq1.4\,{\rm nas}$ and projects to $D_{\rm img}\simeq0.6\,{\rm m}$ at 650 AU, smaller than a metre-class receiver.  The appropriate SGL product is therefore
not a surface image but high-SNR optical/UV spectrophotometry, phase-resolved polarimetry, or pulse timing, potentially constraining atmospheric composition, temperature distribution, magnetospheric emission, or a gravitationally redshifted
line if one is present \cite{Walter1996,OzelFreire2016}.  Quiescent black-hole X-ray binaries occupy a similar regime: the donor star and inner optical-emitting region are usually effectively point-like for SGL imaging, but the amplified mode
could support radial-velocity spectroscopy, fast optical reverberation, and jet polarimetry of otherwise faint systems \cite{RemillardMcClintock2006,CasaresJonker2014}.
These applications are special or opportunistic rather than line-driving, and X-ray SGL operation would require a separate wave-optical and detector analysis.

\subsection{Accretion disks, jets, and active galactic nuclei}

AGN are among the most important non-exoplanet SGL targets because the physical scales of optical/UV accretion disks, BLR gas, and jet bases project to manageable image-plane sizes while the sources are bright \cite{Morgan2010,JimenezVicente2014,Gravity2018BLR,GravityAGN2022}.  The SGL could turn reverberation mapping from a one-dimensional time-delay problem into velocity-resolved spatial tomography, mapping which parts of the BLR respond at which velocities and lags.  It could also test disk temperature profiles, inhomogeneous accretion, disk winds, dust sublimation fronts, and jet-launching structures that are otherwise inferred only indirectly.

AGN provide several compact high-value targets.  Broad-line regions (BLR) in nearby bright AGN can have angular sizes of tens to hundreds of microarcseconds.  Direct near-IR interferometric measurements with GRAVITY have already spatially resolved BLR kinematics in bright quasars, demonstrating the scientific value of velocity-resolved angular information \cite{Gravity2018BLR,GravityAGN2022}.  The SGL could extend this from spectro-astrometric or sparse interferometric constraints to spatially resolved velocity-channel maps, if the line emission is compact enough and the source variability is modeled.

For an AGN BLR of $80\,\uas$, the SGL image-plane diameter is $38\,\km$.  Velocity-resolved imaging would directly constrain the BLR geometry, disk-wind opening angle, inclination, black-hole mass, line responsivity, and outflow/inflow structure.  A compact optical/UV AGN continuum disk provides a complementary scale: a $\sim10^3\,\AU$ disk at $1\,\mathrm{Gpc}$ subtends of order microarcseconds, projects to a kilometre-scale SGL image, and has AU-scale source-plane sampling for metre-scale image-plane motion.  This is the regime in which the SGL could turn microlensing and reverberation constraints into direct spatial or spatial-spectral inference \cite{BlandfordMcKee1982,Peterson1993,Morgan2010,JimenezVicente2014}.  Inner dusty tori are larger and may require sparse scans or selected subregions.  The strongest AGN products are likely parameter inferences from spectrally resolved SGL measurements rather than broadband images.  This is a high-value long-term case.

A related operational advantage is that jet studies can be treated as intra-source campaigns rather than wide-angle retargeting: at 650 AU, stepping by $1''$ along a known jet corresponds to $4.7\times10^5\,{\rm km}\simeq3\times10^{-3}\,{\rm AU}$ in
the image plane, so a focal line selected for a bright AGN core could also support a slow, multi-year sequence of jet-base and downstream knots if the long-wavelength instrument and dynamic calibration requirements are met \cite{AsadaNakamura2012}.

\subsection{Compact line emitters, masers, and high-brightness knots}

Compact line emitters are an important additional target class because the SGL can combine angular resolution with narrowband photon concentration.  Examples include compact molecular-line knots in star-forming regions, submillimetre maser-like emission, circumnuclear line features, and possibly high-brightness megamaser components if longer-wavelength operation is made practical.  The unique product would be a spatially resolved velocity field or line-ratio map at angular scales far below ALMA, ngVLA, or VLBI imaging in the same band.  For star-formation physics this could localize shocks, outflows, and disk winds; for galactic nuclei it could constrain disk geometry, black-hole mass, and warped molecular structures.

This case is wavelength-limited rather than conceptually weak.  Centimetre-wave SGL use is strongly affected by solar plasma and is outside the baseline optical/NIR architecture.  Submillimetre and millimetre line work is more plausible but requires cryogenic receivers, wavelength-specific occultation, and a solar-background model.  These targets should therefore be treated as long-term or speculative high-value extensions of the SGL observatory concept rather than first-generation optical targets.

\subsection{Gravitational-lens systems}

The SGL should be used on gravitational lenses as a follow-up microscope, not as a lens survey.  Conventional surveys and adaptive-optics or space imaging would identify and model the lens; the SGL would then scan a compact lensed image, quasar disk, caustic-crossing star, or bright knot.  The unique products are surface-brightness residuals at extremely small angular scales, microlensing-caustic structure, source-size measurements, and subhalo or line-of-sight perturbations that are washed out by conventional PSFs.

Strong gravitational lenses illustrate the difference between compact components and whole systems.  Macro-image separations of $0.1''$--$5''$ project to $5\times10^4$--$2\times10^6\,\km$, too large for full SGL imaging.  Compact lensed sources are more favorable: quasar accretion disks, lensed stars, microlensing caustic crossings, and compact knots can project to metres or kilometres.  An SGL mission could in principle resolve an individual lensed quasar image or a predicted lensed transient reappearance, constraining accretion-disk size, temperature profile, microlens caustics, or dark substructure; this is directly connected to lensing tests of subhalos and line-of-sight structure \cite{DalalKochanek2002,VegettiKoopmans2009,Gilman2020}.

The inverse problem is complex because the solar lens, foreground lens, microlenses, time delays, source variability, host-galaxy light, and detector/background calibration all enter the measurement.  This case is scientifically valuable but speculative.  It should be pursued only for exceptional preselected systems with strong conventional-lens models and predictable temporal behavior \cite{Wambsganss2006,Kochanek2004}.

\subsection{Nearby galaxies and resolved stellar populations}

The most credible nearby-galaxy use is not galaxy imaging but isolated luminous-star physics.  Cepheids, red supergiants, luminous AGB stars, and extreme variables can have image-plane diameters of metres to kilometres and high intrinsic luminosities.  The SGL could measure limb darkening, pulsation geometry, shock fronts, convection-cell morphology, and dust/asymmetric mass loss in individual stars outside the Milky Way, attacking distance-ladder and late-stage stellar-evolution systematics directly.

The SGL is not a Local Group survey telescope.  Full nearby galaxies subtend arcminutes to degrees and are impossible in image-plane size.  Individual luminous stars are different.  A $1000R_\odot$ red supergiant in M31 has an angular diameter of order $10\,\uas$ and projects to several kilometres.  In principle, the SGL could resolve the surface of an individual star in another galaxy, mapping convection cells, limb darkening, mass loss, and dust asymmetry.

The limitations are crowding, host-galaxy background, target astrometry, and source isolation.  Normal main-sequence stars in M31 are too faint and too compact for useful imaging.  Isolated red supergiants, luminous AGB stars, and selected Cepheids are the plausible targets.  Cepheids are especially interesting because phase-resolved radius variation, limb darkening, shocks, and atmospheric asymmetry are distance-ladder systematics rather than merely stellar-structure details.  This is speculative but potentially revolutionary for standard-candle calibration and massive-star evolution \cite{Riess2022}.

\subsection{Transient phenomena}

The narrow-field nature of the SGL does not eliminate time-domain science; it changes it into a prepositioned or predicted-event program.  The most realistic targets are repeating or forecastable systems: lensed supernova reappearances, recurrent novae, AGN flares, periodic variables, X-ray binaries, and known compact-object systems.  The SGL would provide spatially resolved expansion, asymmetry, line-velocity maps, or evolving compact-source morphology if the focal line is already occupied.

A simple scale check illustrates the opportunity and the limitation.  A photosphere expanding at $10^4\,{\rm km\,s^{-1}}$ reaches a diameter of $\sim350\,{\rm AU}$ after 30 days; at 10 Mpc this corresponds to $\Theta\sim35\,\mu{\rm as}$ and
$D_{\rm img}\sim16\,{\rm km}$, a compact SGL imaging scale.  The obstacle is not angular size or photon rate, but prediction and access: random supernovae are poor targets, whereas lensed-supernova reappearances, recurrent novae, and monitored quasi-periodic systems can convert time-domain SGL astronomy into a prepositioning problem \cite{WangWheeler2008,Kelly2015,Kelly2016}.

Random transients are poor SGL targets because the spacecraft cannot rapidly retarget.  A supernova photosphere or kilonova ejecta could have an attractive angular size, but the focal line must already be occupied.  Plausible transient applications are therefore limited to predictable or preselected events: lensed supernova reappearances, recurrent novae, periodic AGN flares, monitored tidal-disruption-event hosts, or compact-object systems with known electromagnetic phases.  The scientific product would be time-resolved morphology, expansion asymmetry, line-velocity mapping, or microlensing-resolved source evolution.  This is mostly speculative and requires prepositioning or a multi-target architecture over decades.

\subsection{Cosmological structure and high-redshift compact targets}

For cosmology, the SGL is most compelling for compact high-redshift components rather than diffuse galaxies.  Bright quasar accretion disks, lensed stars, super star clusters, nuclear star-forming knots, and caustic events can be treated as selected compact sources.  The unique return would be source sizes, temperature gradients, microlensed structure, and stellar-population or accretion diagnostics at angular scales unavailable to JWST, Roman, ELTs, or natural lensing alone.

A timely example is the population of compact red broad-line sources discovered by JWST, often called ``little red dots'' (LRDs) \cite{Labbe2023,Matthee2024,Greene2024,Kokorev2024}.
For an object at $z\simeq6$, the full $\lesssim100\,{\rm pc}$ system can project to thousands of kilometres in the SGL image plane and is therefore a sparse or long-campaign target.  The compact nucleus or broad-line region is more favorable:
a $\sim10\,{\rm pc}$ nuclear component projects to order $10^3\,{\rm km}$, while a sub-parsec broad-line region can project to kilometre scales.  The credible first SGL mode is amplified thermal-IR spectroscopy or a compact nuclear raster, not
whole-galaxy imaging.  LRDs should therefore be treated as speculative piggyback targets whose feasibility depends on thermal-IR SGL instrumentation, solar-thermal background control, and very stable background subtraction.

The formal linear resolution of the SGL at cosmological distances is striking, but full high-redshift galaxies are not generally feasible.  At angular-diameter distances of order a gigaparsec, parsec-scale regions can subtend $\sim0.1$--$1\,\mas$, which projects to tens to hundreds of kilometres, while hundred-parsec galaxies project to thousands to tens of thousands of kilometres.  Surface-brightness dimming and background contamination are severe.  The plausible cosmological targets are compact AGN, strongly lensed stars, super star clusters, nuclear star-forming knots, and compact caustic events.  These cases are high value but speculative and should be treated as special-target opportunities rather than survey cosmology.

\subsection{SGL transfer-function characterization and solar-system physics}

The SGL transfer function is itself a measurable relativistic optical system, but it should not be treated as an astronomical target class.  Solar quadrupole and higher multipoles, plasma phase, solar oblateness, solar rotation, ephemeris errors, and the telescope/occulter response all imprint chromatic and azimuthal structure on the Einstein ring.  A calibration program that measures ring centroid, azimuthal intensity, wavelength dependence, aperture-averaged gain, polarization response, and temporal variation has two roles.  First, it supplies the transfer-function information required for high-fidelity SGL astronomy.  Second, it is a precision solar-system and relativistic-optics experiment in its own right.  Thus ``SGL transfer calibration'' is better understood as transfer-function characterization: an enabling science product and mission requirement, not a separate imaging target and not evidence for SGL imaging by itself. Therefore, this application is not a separate target class but an enabling science product that should be included in mission requirements and ranking \cite{TuryshevToth2021ExtendedSun,TuryshevToth2023Faint}.

\section{Comparison with foreseeable observatories}
\label{sec:comparison}

Table~\ref{tab:facility_comparison} summarizes the comparison.  The SGL is not simply a larger telescope.  Its advantages are angular resolution, gain for compact targets, and potential spatially resolved spectroscopy at otherwise inaccessible scales.  Its disadvantages are target access, narrow effective field, solar foregrounds, image-plane scanning, PSF complexity, and inverse reconstruction.

Space coronagraphs and starshades are essential for exoplanet detection and spectroscopy, but their angular resolution remains set by telescope diameter.  ELTs provide large collecting area and adaptive-optics imaging but remain at milliarcsecond scales in the near-IR.  CHARA-like optical interferometers reach sub-milliarcsecond resolution for bright stars, but not nanoarcsecond imaging.  ALMA and ngVLA provide high-fidelity millimetre imaging of disks and jets, but not optical/NIR stellar surfaces or white dwarfs.  EHT/ngEHT provide horizon-scale VLBI at tens of microarcseconds, but the SGL can in principle probe sub-microarcsecond structure if long-wavelength SGL operation becomes practical.

The strongest early SGL domain can be summarized as
\begin{equation}
  \theta_{\rm delivered}\lesssim1\,\uas,
  \qquad
  D_{\rm img}\lesssim10^3\,\km,
  \qquad
  \chi_t\lesssim1\ \hbox{or dynamic inversion feasible},
  \label{eq:unique_domain}
\end{equation}
for a target whose flux and background allow calibrated annular measurements.  This domain is not a hard limit on extended-target science; rather, it is the region where a first-generation or modest multi-spacecraft SGL observatory has the clearest advantage.

\begin{table*}[t]
\caption{Comparison of the SGL with other observatory classes. The entries are tied to the resolution and field-access scalings in Figs.~\ref{fig:facility_resolution} and \ref{fig:phase_space}.}
\label{tab:facility_comparison}
\centering
\renewcommand{\arraystretch}{1.10}
\begin{tabular}{lllll}
\toprule
Facility & Main strength & Main limit & SGL contrast & Overlap \\
\midrule
Space coronagraph & \shortstack[l]{Stable high-contrast\\ imaging} & \shortstack[l]{Diffraction-limited\\ resolution} & \shortstack[l]{Sub-$\uas$ mapping\\ of compact targets} & \shortstack[l]{Exoplanets, compact\\ circumstellar structure} \\
Starshade & \shortstack[l]{Broadband external\\ suppression} & \shortstack[l]{No intrinsic\\ resolution gain} & \shortstack[l]{Gain plus extreme\\ angular resolution} & \shortstack[l]{Compact sources,\\ selected disk targets} \\
ELT & \shortstack[l]{Large collecting\\ area} & \shortstack[l]{Atmosphere and\\ mas-scale diffraction} & \shortstack[l]{Access to the\\ nas--$\uas$ regime} & \shortstack[l]{Stars, AGN, bright\\ circumstellar structure} \\
Optical/IR interferometer & \shortstack[l]{Sub-mas\\ imaging} & \shortstack[l]{Sparse $(u,v)$ coverage,\\ limited baselines} & \shortstack[l]{Much finer stellar\\ and WD mapping} & \shortstack[l]{Stars, BLR-related\\ work} \\
ALMA/ngVLA & \shortstack[l]{High-fidelity\\ mm imaging} & \shortstack[l]{Finite\\ baselines} & \shortstack[l]{Sub-AU compact-\\ subfield access} & \shortstack[l]{Disks, jets,\\ compact mm sources} \\
EHT/ngEHT & \shortstack[l]{Horizon-scale\\ VLBI} & \shortstack[l]{Resolution still\\ tens of $\uas$} & \shortstack[l]{Sub-$\uas$ ring and\\ jet substructure} & \shortstack[l]{M87*, compact\\ AGN cores} \\
SGL & \shortstack[l]{Extreme resolution\\ with gain} & \shortstack[l]{Access, solar\\ foregrounds, inversion} & \shortstack[l]{Unique compact-\\ target regime} & \shortstack[l]{White dwarfs, stellar\\ surfaces, AGN} \\
\bottomrule
\end{tabular}
\renewcommand{\arraystretch}{1.0}
\end{table*}

\section{Mission and observing requirements}
\label{sec:mission}

\subsection{Target selection and focal-line architecture}

Because retargeting is extremely expensive, target selection is a mission-level decision.  A general SGL observatory should not be described as a telescope that can point anywhere; it should be described as a focal-line mission with one primary target, a small line-of-sight target set, or a decades-long transfer plan.  Target selection should be based on quantitative closure tests: projected image-plane size, flux/background ratio, variability time, prior ephemeris, scientific uniqueness, and model-inference value.

\subsection{Sampling, navigation, and metrology}

The image-plane raster pitch is
\begin{equation}
  \Delta_{\rm img}=\frac{D_{\rm img}}{n},
  \label{eq:mission_pitch}
\end{equation}
where $n$ is the linear raster dimension and $N=n^2$ is the total number of image-plane samples.  Navigation errors should satisfy
\begin{equation}
  \sigma_\rho\lesssim \epsilon\Delta_{\rm img},
  \label{eq:nav_req}
\end{equation}
with $\epsilon\sim0.01$--$0.05$ for high-fidelity imaging.  For the white-dwarf benchmark, $\Delta_{\rm img}=31\,\m$, so decimeter-to-meter relative metrology is relevant; for a stellar-surface map with kilometer-scale pitch, meter-level metrology is less severe.  For the white-dwarf benchmark this implies decimeter-to-meter \emph{knowledge} of the image-plane coordinate at $650\,\AU$; whether any proposed formation-metrology or ring-centroiding architecture approaches this at the relevant cadence is an open requirement, deferred to the closed-loop navigation simulations listed in Sec.~\ref{sec:limitations}, item~(vi). The measured coordinates, not a post-hoc correction, must enter the forward operator of Eq.~(\ref{eq:joint_inverse}). A zero-mean coordinate error with variance $\sigma_\rho^2$ multiplies the scalar transfer function by the approximate jitter factor
\begin{equation}
  T_{\rm jit}(q)\simeq \exp\!\left[-2\pi^2q^2\sigma_\rho^2\right],
  \label{eq:jitter_otf}
\end{equation}
where $q$ is image-plane spatial frequency. Equation~(\ref{eq:jitter_otf}) is why the requirement is naturally expressed as a fraction of the raster pitch rather than as an absolute distance. The measured spacecraft coordinates should enter the forward operator rather than be corrected after reconstruction.

\subsection{Scan overhead, transverse motion, detector limits, and data volume}
\label{subsec:scan_overhead}

Dwell time alone is not an observing budget.  For an $n\times n$ raster with $N=n^2$ image-plane samples, per-sample dwell time $t_{\rm samp}$, slew/settle time $t_{\rm set}$, calibration overhead $t_{\rm cal}$ per sample or per sample-equivalent block, number of repeat visits $N_{\rm rep}$, and $N_{\rm sc}$ spacecraft, a first-order wall time is
\begin{equation}
  T_{\rm wall}\simeq\frac{N N_{\rm rep}}{N_{\rm sc}}
  \left(t_{\rm samp}+t_{\rm set}+t_{\rm cal}\right)
  =
  \frac{n^2N_{\rm rep}}{N_{\rm sc}}
  \left(t_{\rm samp}+t_{\rm set}+t_{\rm cal}\right).
  \label{eq:wall_time}
\end{equation}
Here $t_{\rm samp}$ is the accumulated dwell assigned to one image-plane sample in one visit; shorter detector exposures or subexposures may be coadded to form that dwell.  For a boustrophedon raster, the total path length per complete map is approximately $L_{\rm scan}\simeq nD_{\rm img}$, and the per-spacecraft path length is $L_{\rm scan}/N_{\rm sc}$ if the raster is divided evenly.  Continuous scanning must also satisfy a smear condition
\begin{equation}
  v_\perp t_{\rm exp}\lesssim f_{\rm smear}\Delta_{\rm img},
  \label{eq:smear}
\end{equation}
where $t_{\rm exp}\le t_{\rm samp}$ is the individual exposure or subexposure time relevant for motion smear. For step-and-integrate observing, the corresponding requirement is slew/settle repeatability and accurate knowledge of the actual sample coordinates.

\begin{table*}[t]
\caption{First-order operational scales implied by the scalar benchmarks before repeated phase coverage, slew/settle overhead, calibration blocks, and downlink.  The raster convention is $n$ for the linear dimension and $N=n^2$ for the total number of image-plane samples.  The ideal integrated dwell is $T_{\rm dwell}=Nt_{\rm samp}/N_{\rm sc}=n^2t_{\rm samp}/N_{\rm sc}$.  The single-map raster path length is $L_{\rm scan}\simeq nD_{\rm img}$.  The quantity $\Delta_{\rm img}/t_{\rm exp}$ is not a required velocity; it is the velocity corresponding to one image-plane pitch per exposure and therefore an upper diagnostic scale for continuous scanning.}
\label{tab:scan_budget}
\centering
\renewcommand{\arraystretch}{1.08}
\begin{tabular}{lcccc}
\toprule
Case & $T_{\rm dwell}$ & $L_{\rm scan}$ & $\Delta_{\rm img}/t_{\rm exp}$ & Main operational issue \\
 & [days] & [km] & [m s$^{-1}$] & \\
\midrule
Solar analog, $10\,\pc$ & 0.074 & $7.0\times10^4$ & $1.37\times10^3$ & fast scan or many repeats; variability \\
Magnetic white dwarf & 0.95 & $5.1\times10^2$ & 1.6 & metrology, polarimetry, spin phase \\
M87*-like ring/jet & 0.47 & $2.5\times10^3$ & 5.2 & mm receiver, polarization, dynamics \\
$0.1\,\AU$ disk subfield & 4.7 & $4.3\times10^4$ & 8.8 & background, host leakage, long campaign \\
\bottomrule
\end{tabular}
\renewcommand{\arraystretch}{1.0}
\end{table*}

Table~\ref{tab:scan_budget} summarizes first-order dwell, scan-path, and continuous-scan velocity scales for the scalar benchmark cases before repeated phase coverage, slew/settle overhead, calibration blocks, and downlink.

Detector limits are also target dependent. Bright stellar and white-dwarf cases can have annular source rates high enough that saturation, full-well management, detector linearity, and ring-sector dynamic range matter more than photon statistics. Conversely, selected disk subfields and reflected-light exoplanets remain background and calibration limited. If the ring is compressed to $N_a$ azimuthal sectors, $N_\lambda$ spectral channels, $N_p$ polarization states, and $b_{\rm samp}$ bytes per integrated sample, the compressed science data volume per raster is approximately
\begin{equation}
  V_{\rm sci}\simeq N N_aN_\lambda N_p b_{\rm samp}N_{\rm rep}
  = n^2N_aN_\lambda N_p b_{\rm samp}N_{\rm rep}.
  \label{eq:data_volume}
\end{equation}
For $n=160$ ($N=2.56\times10^4$), $N_a=32$, $N_\lambda=300$, $N_p=4$, and $b_{\rm samp}=4$ bytes, this is $\simeq3.9\,\mathrm{GB}$ per compressed raster before housekeeping and calibration products. Raw detector frames can be much larger, so onboard ring extraction, compression, autonomy, and optical communication are observatory requirements rather than mission-support details. Multiple spacecraft or a distributed focal-shell architecture reduce wall time and improve simultaneity, but they also introduce formation-control, inter-spacecraft metrology, clock synchronization, cross-calibration, and relay/downlink requirements.

\subsection{Solar suppression, corona, and detector calibration}

The annular measurement occurs close to the solar limb, so the mean corona, coronal shot noise, solar thermal leakage, stray light, detector response, and occulting architecture are first-order science requirements.  Multiplicative residuals on a bright background appear in source units as
\begin{equation}
  \delta_{\rm src}=\epsilon_{\rm cal}\frac{Q_b}{Q_s}.
  \label{eq:false_signal}
\end{equation}
Thus the calibration requirement depends on the target class.  Bright stellar targets can tolerate larger fractional background residuals than faint protoplanetary subfields, while exoplanets and diffuse sources can require sub-ppm residuals.  External occulters may be important for broad spectral coverage and for reducing internal coronagraph losses \cite{TuryshevToth2022Spectral}.

\subsection{Power, communication, and distributed operations}

\begin{figure*}[t]
  \centering
  \includegraphics[width=0.86\textwidth]{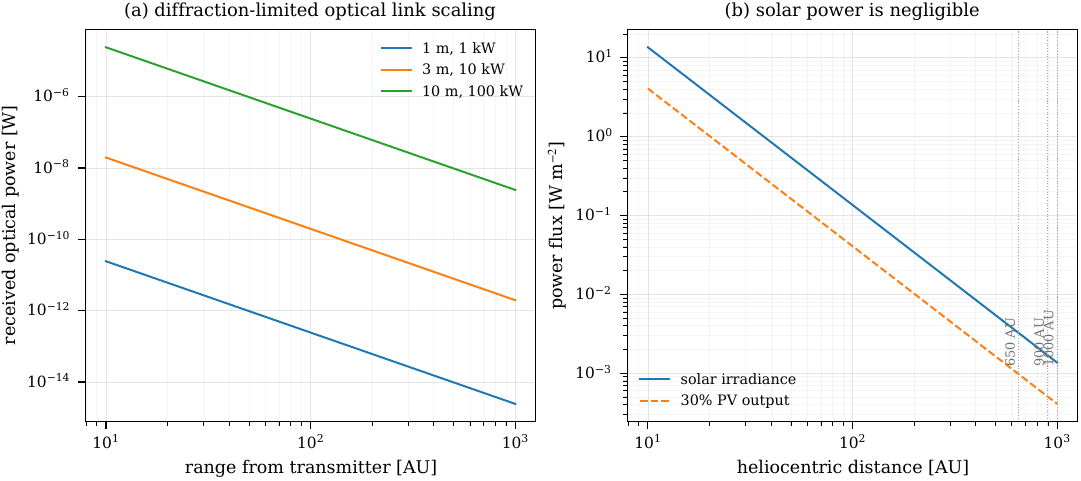}
  \caption{Power and communication constraints for SGL observatory architectures.  Panel (a) gives an approximate diffraction-limited optical-link received-power scaling at $1064\,\mathrm{nm}$ for representative transmitter/receiver apertures and transmitter powers.  Panel (b) shows solar irradiance and a 30\% photovoltaic-output surrogate versus heliocentric distance.  The curves are order-of-magnitude design scalings, but they show why onboard power, optical communications, autonomy, and data compression are first-order observatory requirements at $650$--$1000\,\AU$.}
  \label{fig:power_comms}
\end{figure*}

Solar arrays are not a credible primary power source at SGL distances: the solar irradiance at $650\,\AU$ is only $1361/650^2=3.2\times10^{-3}\,\mathrm{W\,m^{-2}}$.  A serious observatory therefore requires radioisotope, fission, beamed, or more advanced power.  Communications are similarly mission-defining.  A diffraction-limited optical link with transmitter diameter $D_t$, receiver diameter $D_r$, transmit power $P_t$, communication wavelength $\lambda_c$, range $R$, and end-to-end link efficiency $\eta_{\rm link}$ has the approximate received power
\begin{equation}
  P_r\simeq P_t\eta_{\rm link}\Big(\frac{\pi D_tD_r}{4\lambda_cR}\Big)^2 .
  \label{eq:optical_link}
\end{equation}
This scaling makes high-gain optical communication, autonomy, onboard compression, and long-latency operations part of the observatory design rather than mission-support details.  Multiple spacecraft are not required by SGL physics, but they can reduce wall-clock scan time, sample variable sources quasi-simultaneously, separate telescope/occulter/relay functions, and provide redundancy for calibration and navigation.  Local short-baseline interferometry within a formation may help ring extraction and nulling, but it does not replace the solar gravitational aperture; matching the SGL's effective baseline would require a qualitatively different million-kilometre-scale architecture. Figure~\ref{fig:power_comms} illustrates why power, optical comm, autonomy, and compression are first-order observatory requirements at $650$--$1000\,\AU$.

\subsection{Dynamic reconstruction}

The mission data product should be a posterior over physical maps and nuisance parameters, not a single deblurred image.  A general dynamic inverse problem is
\begin{equation}
  \min_{O(t),\bm{\eta}}
  \left\|C_n^{-1/2}\left[y-H(\bm{\eta})O(t)-b(\bm{\eta})\right]\right\|^2
  +R[O(t),\bm{\eta}],
  \label{eq:joint_inverse}
\end{equation}
where $\bm{\eta}$ includes PSF multipoles, pointing, background templates, detector calibration, ephemeris, spectral response, and target-geometry parameters.  For stellar surfaces, $O(t)$ should include rotation and active-region evolution.  For black holes, it should include time-dependent accretion-flow morphology and polarization.  For AGN, it should include velocity-resolved line response.  For protoplanetary subfields, it should include orbital motion, scattered-light phase, and accretion variability.

\section{Prioritized non-exoplanet applications}
\label{sec:ranking}

Table~\ref{tab:ranking} gives a technology-confidence hierarchy.  The ordering is not a yield forecast; it ranks how cleanly each science case closes the variables in Eq.~(\ref{eq:closure_vector}).  White dwarfs are the strongest bounded optical/NIR imaging case.  Nearby stellar surfaces are comparably valuable but require dynamic reconstruction.  Compact black-hole and AGN applications are high-value long-wavelength concepts until receiver, plasma, solar-thermal, occultation, polarization, and dynamic-imaging terms close.  Planet-forming subfields are challenging but important because the decisive spatial scale is compact even when the full disk is not.  Transfer-function characterization is placed as an enabling program because every quantitative SGL image requires a calibrated SGL+telescope response.

\begin{table*}[t]
\caption{Qualitative prioritization of non-exoplanet SGL applications. The tiers combine phase-space closure, scientific value, and implementation maturity. The NS/BH-XRB entry refers to the amplified non-imaging mode defined in Sec.~\ref{sec:phase_space}, not to resolved surface imaging. ``Near'' means plausible for a first-generation optical/NIR observability study if the line of sight is selected; ``long'' means high value but dependent on specialized instrumentation, multiple spacecraft, or mature PSF/dynamic modeling; ``spec.'' means speculative or opportunistic. The SGL transfer-function characterization row is an enabling calibration and solar-system-physics program, not a resolved astronomical target class.}
\label{tab:ranking}
\centering
\renewcommand{\arraystretch}{1.08}
\setlength{\tabcolsep}{4pt}
\begin{tabular*}{\textwidth}{@{\extracolsep{\fill}}llccl}
\toprule
Tier & Application & Maturity & Risk & Controlling limitation \\
\midrule
Highest optical/NIR & WD surface and magnetic maps & near & med. & PSF; metrology; spectropolarimetry \\
Highest optical/NIR & Nearby stellar maps and activity & near & med. & variability; dynamics; detector range \\
Highest enabling & SGL transfer-function  & long & med. & solar multipoles; plasma; extended Sun;  \\
 &  characterization &  & &  ring extraction \\
High-value long-term & BH/AGN compact accretion & long & high & mm receiver; plasma; dynamics \\
High-value long-term & AGN BLR velocity maps & long & high & spectroscopy; reverberation; cadence \\
High-value long-term & Planet-forming subfields & long & high & host leakage; background; tracers \\
High-value long-term & Compact AGN disks & long & high & reverberation; spectra; variability \\
High-value long-term & Evolved-star inner winds & long & med. & envelope size; dust transfer; variability \\
 Special/opportunistic & NS/BH-XRB amplified  & spec. & med.--high & line content; access; dynamic range \\
 & spectroscopy and timing &  &&  \\
Special/opportunistic & Compact lensed sources & spec. & high & lens model; microlensing; delays \\
Special/opportunistic & Compact line/maser knots & spec. & high & long wavelength; plasma; calibration \\
Special/opportunistic & Local Group luminous stars & spec. & high & crowding; background; phase coverage \\
Special/opportunistic & Predictable transients & spec. & high & prepositioning; event prediction \\
Special/opportunistic & High-$z$ compact sources/ & spec. & high & thermal-IR background;  \\
 &  LRD nuclei & &  & surface brightness;  preselection \\
\bottomrule
\end{tabular*}
\setlength{\tabcolsep}{6pt}
\renewcommand{\arraystretch}{1.0}
\normalsize
\end{table*}

The prioritization in Table~\ref{tab:ranking} is a capability assessment, not a yield prediction. A target class may be scientifically compelling and photon-rich while still being operationally expensive if its line of sight is inaccessible from a given trajectory. A future yield analysis must combine sky distribution, target priority, heliocentric transfer time, transverse retargeting cost, spacecraft lifetime, power, downlink, and the possibility of distributed focal shells or target-line clusters. The narrower question addressed here is which astronomical measurements become uniquely possible once an SGL spacecraft is assigned to a suitable line of sight.

For mission design, the relevant unit is not always a single target but a focal-line portfolio.  Degree-scale retargeting remains prohibitive, but compact sources within a small angular neighborhood of a chosen line may be observed with much smaller image-plane offsets.  A line selected for a nearby white dwarf, stellar target, or exoplanet host could therefore also have secondary value if it contains a background AGN, compact high-redshift source, lensed transient reappearance, recurrent nova, or amplified point-source compact remnant.  Future yield studies should optimize over such target bundles rather than ranking isolated objects only.

\section{Limitations and next steps}
\label{sec:limitations}

The results in this paper are traceable but deliberately limited. They use scalar aperture-averaged SGL simulations, not full focal-plane diffraction models. The main four-image gallery uses generic anisotropic PSF stress-test parameters rather than a physical solar $J_2$, $J_4$, spin-axis, plasma, or extended-Sun propagation. The added white-dwarf $J_2/J_4$ closure test is a useful intermediate step because it ties the residual kernel error to a solar-multipole image-plane scale, but it is still a scalar displacement scale rather than a wave-optical solar PSF library. The SSIM-versus-$\SNRc$ sweep added in Fig.~\ref{fig:sensitivity_boundaries}(b) makes the information-floor dependence explicit, but it does not derive that floor from a detector, coronal-background, or calibration-covariance model. The optical/NIR effective gain is a finite-source-limited planning model, not a final annular-instrument throughput. The M87*-like case uses demonstration millimetre gain and background assumptions, not a detailed receiver, occultation, solar thermal, plasma, and calibration model. The synthetic target scenes are physically motivated analytic benchmarks, not stellar-MHD, white-dwarf atmosphere and spectropolarimetric, GRMHD, AGN BLR, disk radiative-transfer, or lensing scene libraries. These limitations are appropriate for a phase-space and benchmark-reconstruction paper, but they define the boundary between the present work and mission validation.

The remaining technology risks are finite and measurable.  Table~\ref{tab:validation_status} separates what is demonstrated in the present benchmark from what must be verified before target-specific observatory performance can be claimed.  The benchmark result is not that these closure variables vanish; it is that the dominant variables have been isolated, scaled, and tied to quantitative diagnostics.

\begin{table*}[t]
\caption{Validation status of the load-bearing reductions in this paper.  The entries distinguish demonstrated scalar recoverability from the technology products needed for target-specific SGL observatory performance.}
\label{tab:validation_status}
\centering
\small
\renewcommand{\arraystretch}{1.15}
\setlength{\tabcolsep}{4pt}
\begingroup
\def\TableXIcell#1#2{%
  \parbox[t]{#1}{%
    \setlength{\parindent}{0pt}%
    \leftskip=0pt\relax
    \rightskip=0pt\relax
    \parfillskip=0pt plus 1fil\relax
    #2%
  }%
}
\begin{tabular}{llll}
\toprule
\TableXIcell{0.09\textwidth}{Closure variable}
&
\TableXIcell{0.22\textwidth}{What is demonstrated here}
&
\TableXIcell{0.31\textwidth}{Quantitative confidence from this paper}
&
\TableXIcell{0.32\textwidth}{Verification product needed}
\\
\midrule

\TableXIcell{0.09\textwidth}{Annular/ vector operator}
&
\TableXIcell{0.22\textwidth}{The scalar aperture-averaged inverse preserves selected morphologies afterxnvolution, structured residuals, support constraints, and deconvolution.}
&
\TableXIcell{0.31\textwidth}{Four fixed-protocol benchmarks give $\SSIM=0.993$, $0.918$, $0.973$, and $0.923$, with $\FRC_{50}$ and leakage diagnostics in Table~\ref{tab:metrics}.}
&
\TableXIcell{0.32\textwidth}{Propagate the same target classes through Eq.~(\ref{eq:vector_measurement}) with wavelength, ring sector, polarization, detector covariance, occultation, and fitted nuisance parameters.}
\\

\TableXIcell{0.09\textwidth}{Effective $\SNRc$ floor}
&
\TableXIcell{0.20\textwidth}{The adopted $\SNRc$ values are explicit information-quality floors, not photon-only limits.}
&
\TableXIcell{0.31\textwidth}{Figure~\ref{fig:sensitivity_boundaries}(b) gives $\SSIM(\SNRc)$ with scene, support, kernel family, and regularization fixed.  For the white dwarf, $\SNRc^{\rm phot}\sim10^4$--$10^5$ per sample while $\SNRc^{\rm eff}=95$ is used.}
&
\TableXIcell{0.32\textwidth}{Derive $\SNRc^{\rm eff}$ from detector photon-transfer data, coronal/thermal foregrounds, annular extraction, calibration covariance, metrology error, and dynamic source residuals.}
\\

\TableXIcell{0.09\textwidth}{Solar transfer function}
&
\TableXIcell{0.22\textwidth}{Generic anisotropic kernels and the white-dwarf $J_2/J_4$ residual test quantify inverse sensitivity to deterministic PSF error.}
&
\TableXIcell{0.31\textwidth}{A residual $f_J=0.05$ gives $14\,\m=0.45\Delta_{\rm img}$ and preserves the main morphology; $f_J=1$ gives $281\,\m\simeq9\Delta_{\rm img}$ and becomes a controlling error.}
&
\TableXIcell{0.32\textwidth}{Build a wavelength-dependent solar-multipole, plasma, extended-Sun, annular-extraction, detector, and metrology transfer-function library and demonstrate residuals below the science-contrast or fractional-pitch requirement.}
\\

\TableXIcell{0.09\textwidth}{M87*/mm application}
&
\TableXIcell{0.22\textwidth}{Compact black-hole structure is geometrically well matched to the SGL image plane and recoverable at the stipulated scalar information quality.}
&
\TableXIcell{0.31\textwidth}{The M87*-like row uses $D_{\rm img}=19.8\,\km$, $G_{\rm eff}=10^4$, $Q_b=3.0\times10^{10}\,\mathrm{s^{-1}}$, $\SNRc=38$, and recovers $\FRC_{50}=0.66\,\mu$as.}
&
\TableXIcell{0.32\textwidth}{Demonstrate a millimetre/sub-millimetre SGL receiver model including solar thermal emission, plasma phase, scattering, occultation, annular coupling, polarization leakage, and dynamic imaging.}
\\

\TableXIcell{0.09\textwidth}{Metric stability}
&
\TableXIcell{0.22\textwidth}{The reported metrics are fixed-protocol scalar diagnostics, not ensemble performance claims.}
&
\TableXIcell{0.31\textwidth}{Table~\ref{tab:metrics} reports one fixed-seed realization per case; deterministic variations from PSF residuals, masks, floors, and regularization are already larger than the random draw for the intended use.}
&
\TableXIcell{0.32\textwidth}{Promote the benchmarks to ensemble posteriors over seeds, PSF orientation, physical solar geometry, calibration residuals, support priors, raster pitch, dwell, cadence, and regularization.}
\\
\bottomrule
\end{tabular}
\endgroup
\setlength{\tabcolsep}{6pt}
\renewcommand{\arraystretch}{1.0}
\normalsize
\end{table*}

The next development stage should therefore convert each entry of Table~\ref{tab:validation_status} from an assumed or surrogate quantity into a measured covariance term in the full SGL observing model.  The required work items are:
 (i) a solar multipole and plasma wave-optical PSF library that supersedes the $J_2/J_4$ scalar displacement scale used here; (ii) full annular instrument, occulting, detector, and ring-extraction propagation; (iii) detector photon-transfer and calibration covariance models; (iv) astrophysical scene libraries for stellar surfaces, white dwarfs, GRMHD black-hole images, AGN BLR models, disk radiative-transfer subfields, and compact lensed sources; (v) dynamic Bayesian reconstruction with source and nuisance-parameter posteriors; (vi) closed-loop image-plane navigation and metrology simulations; (vii) target-specific sensitivity curves beyond the scalar $\SNRc$ and $J_2/J_4$ sweeps shown here, including PSF mismatch, calibration residuals, background errors, gain assumptions, metrology, pointing, raster pitch, dwell time, and temporal variability; and (viii) target-specific yield studies that include focal-line access, cruise architecture, power, downlink, and distributed-observatory options. The guiding requirement is that every science claim be tied to target angular size, image-plane size, photon rate, background ratio, scan time, variability, PSF model, reconstruction metric, and uniqueness relative to alternative facilities.

\section{Conclusions}
\label{sec:conclusion}

We have shown that the SGL is a credible ultra-high-resolution observability platform for selected non-exoplanet targets once it is treated as a calibrated focal-line inverse problem rather than as a conventional telescope.  The results identify where the physics is favorable, what scalar recovery demonstrates, and which technology variables control delivered observatory performance.  The main conclusions are as follows.

First, the SGL is a constrained gravitational observatory, not a general survey telescope. Its formal gain and angular response are extraordinary, but the operational science is determined by image-plane sampling, target access, solar foregrounds, PSF knowledge, source variability, and inverse reconstruction.

Second, the most important feasibility scaling is the vector mapping $\boldsymbol{\rho}=-z\boldsymbol{\theta}=-(z/z_0)\boldsymbol{\xi}$, whose scalar diameter is $D_{\rm img}=z\Theta$ and whose image-plane sampling pitch is $\Delta_{\rm img}=D_{\rm img}/n$, with $N=n^2$ total image-plane samples.  At $650\,\AU$, $1\,\uas$ corresponds to $0.471\,\km$ and $1''$ to $4.71\times10^5\,\km$; the finite-source gain scale further obeys $\mu_{\rm ext}D_{\rm img}\simeq4b$.  This favors compact targets and selected subfields while disfavoring whole disks, whole lens systems, and full galaxies.

Third, the scalar benchmarks demonstrate useful recoverability for four representative target regimes.  A solar analog at $10\,\pc$, a magnetic white dwarf at $10\,\pc$, an M87*-like ring/jet-base source, and a $0.1\,\AU$ protoplanetary subfield all retain scientifically meaningful spatial information after SGL scalar convolution and deconvolution under the stated benchmark protocol.  The reported $\SSIM$, $\FRC_{50}$, leakage, contrast, and $\SNRr$ values establish the favorable inverse-conditioning regime for the adopted kernels, masks, effective $\SNRc$ floors, and regularization.  Delivered performance is then set by closure of the calibrated transfer function $H_{\rm cal}$, annular extraction, detector/background covariance, image-plane metrology, and source dynamics.  The white-dwarf case is the cleanest static benchmark; stellar surfaces and protoplanetary subfields move directly into the dynamic-inversion regime.

Fourth, the strongest non-exoplanet astronomical applications are white-dwarf surface and magnetic mapping, nearby stellar surfaces and activity, compact black-hole and AGN structure requiring dedicated long-wavelength receiver/background modeling, velocity-resolved AGN BLR imaging, and selected planet-forming subfields.  SGL transfer-function characterization is a separate highest-priority enabling program: it calibrates the solar-multipole, plasma, extended-Sun, annular-extraction, detector, and metrology terms needed to interpret any reconstructed SGL image. Strong-lens compact sources, compact line/maser-emission knots, isolated luminous stars and Cepheids in nearby galaxies, predictable transients, high-$z$ compact sources/LRD nuclei, and relativistic-optics calibration are scientifically interesting but more specialized, more architecture-limited, or more wavelength-dependent.  Not every valuable SGL observation must be a reconstructed image: for targets with $D_{\rm img}\lesssim d$ or $D_{\rm img}\lesssim\Delta_{\rm img}$, amplified photometry, spectroscopy, polarimetry, astrometry, reverberation mapping, or time-domain monitoring may be the scientifically appropriate data product.

Finally, the technology path is well defined.  A quantitative SGL observatory must deliver calibrated ring extraction, solar-corona and solar-thermal suppression, target-appropriate background residuals, physical solar-multipole and plasma transfer functions, detector calibration covariance, image-plane metrology, dynamic reconstruction, onboard data reduction, and preselected focal-line operations.  These requirements are demanding but specific.  The key distinction from reflected-light exoplanet imaging is that many non-exoplanet targets are photon-rich, so the limiting variables become calibration, transfer-function knowledge, dynamic range, temporal modeling, and access geometry.  If those variables close at the levels quantified here, the SGL enables astronomical measurements at angular scales unavailable to foreseeable telescopes, interferometers, coronagraphs, starshades, or VLBI arrays.

\section*{Acknowledgments}

The author thanks Alexander Kusenko for stimulating discussions, encouragement, and  valuable comments while this document was in preparation. The work described here was carried out at the Jet Propulsion Laboratory, California Institute of Technology, Pasadena, California, under a contract with the National Aeronautics and Space Administration.  
\textcopyright\ 2026. California Institute of Technology. Government sponsorship acknowledged.

\appendix

\section{Benchmark implementation details for scalar simulations}
\label{app:benchmark_details}

This appendix records the  conventions used for the scalar benchmark simulations.  The purpose is to make the assumptions, normalizations, kernel construction, noise model, reconstruction procedure,  metric definitions explicit within the paper.  These details define the controlled observability benchmarks; mission-level validation would require replacing the analytic scenes/scalar kernels with target-specific astrophysical scene libraries, a physical SGL+telescope PSF and ring-response library, detector-calibration covariance, closed-loop metrology,  dynamic reconstruction.

\subsection{Common grid, mapping, and normalization}

Each source scene is represented on an $n\times n$ Cartesian grid in the SGL image plane, with total sample count $N=n^2$.  The mapping from source-plane coordinate $\boldsymbol{\xi}$ to image-plane coordinate $\boldsymbol{\rho}$ is
\begin{equation}
  \boldsymbol{\rho}=-\frac{z}{z_0}\boldsymbol{\xi},
  \label{eq:app_mapping}
\end{equation}
with scalar image-plane diameter $D_{\rm img}=z\Theta$ and pitch $\Delta_{\rm img}=D_{\rm img}/n$.  This is the same physical image-plane sampling quantity denoted $\Delta_{\rm img}$ in the resolved exoplanet-imaging benchmark.  The target distance is denoted $z_0$ throughout the shared SGL geometry, while $z_{\rm foc}$ denotes the solar-grazing focal onset.

The parity inversion is applied when forming the image-plane scene and reversed when reporting source-plane products. The scene array is first normalized to unit mean over the science support mask $M$ and then scaled so that the integrated unlensed photon rate and SGL-coupled source rate match the $Q_s$ value in Table~\ref{tab:sim_assumptions}. Display images may be peak-normalized or stretched for visibility, but metrics are computed before display stretching.

\subsection{Benchmark scenes and preprocessing}

The analytic scenes contain only dimensionless morphology before radiometric scaling. The solar analog uses a limb-darkened disk, a granulation-like random field filtered to suppress pixel-scale noise, dark spot components, and bright plage-like components. The white-dwarf scene uses a limb-darkened compact disk with broad hot and cool regions, an accretion-belt surrogate, and a smooth Zeeman/Stokes magnetic surrogate. The M87*-like scene uses a crescent ring, a central shadow depression, a compact hot spot, and a jet-base extension on a millimetre image-plane scale. The protoplanetary subfield uses a smooth background gradient, a compact circumplanetary-disk surrogate, a spiral wake, a gap-edge gradient, a dust-trap component,  an accretion-shock surrogate. Before convolution, negative pixels are clipped to zero, the support mask is applied,  the image is scaled to the intended source rate. These analytic scenes are benchmarks for information recovery; target-specific validation should replace them with stellar-MHD, white-dwarf atmosphere and spectropolarimetric, GRMHD, AGN BLR, disk radiative-transfer/lensing simulations.

\subsection{Kernel construction and PSF perturbations}

The nominal scalar kernel is the aperture-averaged light-bucket kernel
\begin{equation}
  K_0(0)=1,\qquad K_0(\rho>0)=\frac{d}{4\rho},
\end{equation}
computed on the same image-plane grid, truncated at the computational field, and normalized to conserve total kernel weight on that field. The truth kernel is obtained by applying Eq.~(\ref{eq:sim_psf_mismatch}) and renormalizing the result.

The perturbation orientation $\phi_0$ is fixed for each benchmark definition; mission-level sensitivity studies should repeat the reconstruction over multiple $\phi_0$ values and physical solar geometries.  The reconstruction kernel is always the nominal $K_0$, so the mismatch is not known to the inverse solver.

\subsection{\texorpdfstring{$J_2/J_4$ white-dwarf closure-test settings}{J2/J4 white-dwarf closure-test settings}}

The additional closure test in Fig.~\ref{fig:wd_j_multipole} uses the white-dwarf grid in Table~\ref{tab:sim_assumptions}: $n=128$, $N=n^2=16384$, $D_{\rm img}=4.02\,\km$, $\Delta_{\rm img}=31.4\,\m$, and $\SNRc=95$. The analytic source is a limb-darkened disk with broad hot and cool regions, an accretion-belt surrogate, and a smooth low-order magnetic/temperature asymmetry. It is normalized to unit mean on the disk support. The nominal kernel is $K_0$ above. The truth kernel for a residual fraction $f_J$ is given by Eqs.~(\ref{eq:j_proxy_kernel})--(\ref{eq:j_proxy_scale}) with $\phi_\odot=25^\circ$, $J_2=2.2\times10^{-7}$, $|J_4|=4.0\times10^{-9}$, $R_\odot=6.957\times10^8\,\m$, and $b=1.089R_\odot$. The random seed for the plotted realization is 4268. Gaussian noise is scaled so that the convolved raster has $\SNRc=95$. The reconstruction uses the same Wiener inverse as Eq.~(\ref{eq:wiener}) with the nominal kernel $K_0$ and fixed $\gamma=10^{-2}$, followed by non-negativity, disk masking, and a single masked-mean photometric rescaling. Under this protocol the auxiliary closure metrics are $\SSIM=0.933$, $0.852$, $0.765$, $0.695$, and $0.628$ for $f_J=0$, $0.05$, $0.20$, $0.50$, and $1.0$, respectively; the corresponding NRMSE values are $0.074$, $0.144$, $0.183$, $0.252$, and $0.294$. These values are meant to quantify the residual-multipole sensitivity of the scalar inverse, not to replace the wave-optical PSF library required for mission validation.

\subsection{Noise, background, and calibration floors}

For each benchmark, the scalar measurement is
\begin{equation}
  y_i=(H_{\rm true}O)_i+b_i+n_i+s_i,
\end{equation}
where $b_i$ is the mean residual background, $n_i$ is a Gaussian approximation to photon noise at the adopted source/background rate, and $s_i$ is a low-frequency structured residual field. The variance of $n_i$ and the amplitude of $s_i$ are scaled so that the convolved raster has the target $\SNRc$ in Table~\ref{tab:sim_assumptions}. In the benchmark implementation, $s_i$ may be generated by drawing a unit-variance Gaussian random field in Fourier space, multiplying by a low-pass filter, transforming back to the image plane, subtracting the mean on $M$, and scaling to the desired systematic floor.  The adopted floors represent residual background subtraction, low-order PSF error, detector calibration, and unresolved temporal evolution at the scalar-benchmark level.

\subsection{Effective-SNR sensitivity diagnostic}

The $\SSIM$-versus-$\SNRc$ curves in Fig.~\ref{fig:sensitivity_boundaries}(b) are generated by holding the scalar scenes, masks, kernel family, support priors, and regularization convention fixed while varying only the effective convolved-raster information floor.  The reported benchmark $\SSIM$ values are used as anchor points at the adopted $\SNRc$ values in Table~\ref{tab:sim_assumptions}; the curves then show the response to lower and higher effective floors.  This diagnostic is intentionally narrower than a physical detector/background model.  It separates inverse-problem robustness from the question of whether a flight instrument can deliver a specified $\SNRc$ after solar-background subtraction, detector calibration, PSF modeling, metrology, and source variability.

\subsection{Reconstruction and regularization}

The baseline reconstruction is the Fourier/Wiener inverse in Eq.~(\ref{eq:wiener}) using the nominal kernel $K_0$. The regularization parameter $\gamma$ is selected from the adopted noise level and the kernel transfer function, not from the hidden truth. A deterministic implementation can use a logarithmic grid in $\gamma$, choose the L-curve corner~\cite{Hansen1992Lcurve} or generalized-cross-validation minimum~\cite{GolubHeathWahba1979GCV} using only $y$, $K_0$, and the adopted noise variance, and then apply the same choice to all metric calculations. After inversion, non-negativity is imposed, the support mask is applied for disk-like cases, and a single multiplicative photometric scale factor is fitted on $M$ so that the reconstructed and true maps have the same masked mean flux.

\subsection{Evaluation metrics}

Metrics are computed on the mask $M$ using Eqs.~(\ref{eq:nrmse_def})--(\ref{eq:snrr_def}). SSIM uses the same photometric normalization and dynamic range as NRMSE. FRC$_{50}$ is obtained by Fourier-transforming the masked truth and reconstruction, computing the normalized cross-correlation in annuli of spatial frequency, and converting the first crossing of 0.5 to the natural source unit. For disk sources the conversion uses the source-plane scale $\Delta s=(z_0/z)\Delta_{\rm img}$; for M87* it is reported as an angular scale; for the protoplanetary subfield it is reported in AU.

\section{Recommended robustness sweeps for mission validation}
\label{app:robustness}

A target-specific mission study should repeat the benchmarks over a multi-dimensional grid rather than reporting a single scalar reconstruction. The minimum recommended sweeps are: $\epsilon_2$ and $\epsilon_4$ in Eq.~(\ref{eq:sim_psf_mismatch}); PSF orientation and chromaticity; multiplicative and additive background errors; calibration floor $\epsilon_{\rm cal}$; finite-source gain and selected-field size; Gaussian coordinate jitter and slow pointing drift; linear raster dimension $n$, total sample count $N=n^2$, and pitch $\Delta_{\rm img}$; dwell time $t$; and source evolution with $\chi_t=T_{\rm wall}/t_{\rm var}$. The output should be a success/failure boundary for each science observable, not only a set of visually successful reconstructions.


%

\end{document}